\documentclass[]{aastex631}
\usepackage{float}
\usepackage{soul}
\usepackage[caption = false]{subfig}

\shorttitle{Performance of HAWC with the improved reconstruction algorithms}

\graphicspath{{./}{figures/}}
\begin{document}

\title{Performance of the HAWC Observatory and TeV Gamma-Ray Measurements of the Crab Nebula with Improved Extensive Air Shower Reconstruction Algorithms}

\correspondingauthor{Sohyoun Yun-C\'arcamo}
\email{yunsoh@umd.edu}

\author{A.~Albert}
\affiliation{Physics Division, Los Alamos National Laboratory, Los Alamos, NM, USA }

\author{R.~Alfaro}
\affiliation{Instituto de F\'{i}sica, Universidad Nacional Aut\'onoma de M\'exico, Ciudad de M\'exico, M\'exico }

\author{C.~Alvarez}
\affiliation{Universidad Aut\'onoma de Chiapas, Tuxtla Guti\'errez, Chiapas, M\'exico}

\author{A.~Andr\'es}
\affiliation{Instituto de Astronom\'ia, Universidad Nacional Aut\'onoma de M\'exico, Ciudad de M\'exico, M\'exico }

\author{J.C.~Arteaga-Vel\'azquez}
\affiliation{Universidad Michoacana de San Nicol\'as de Hidalgo, Morelia, M\'exico }

\author{D.~Avila Rojas}
\affiliation{Instituto de F\'isica, Universidad Nacional Aut\'onoma de M\'exico, Ciudad de M\'exico, M\'exico }

\author{H.A.~Ayala Solares}
\affiliation{Department of Physics, Pennsylvania State University, University Park, PA, USA }

\author{R.~Babu}
\affiliation{Department of Physics and Astronomy, Michigan State University, East Lansing, MI, USA }

\author{E.~Belmont-Moreno}
\affiliation{Instituto de F\'isica, Universidad Nacional Aut\'onoma de M\'exico, Ciudad de M\'exico, M\'exico }

\author{A.~Bernal}
\affiliation{Instituto de Astronom'ia, Universidad Nacional Autónoma de México, Ciudad de M\'exico, M\'exico }

\author{K.S.~Caballero-Mora}
\affiliation{Universidad Autónoma de Chiapas, Tuxtla Gutiérrez, Chiapas, México}

\author{T.~Capistr\'an}
\affiliation{Instituto de Astronom\'ia, Universidad Nacional Aut\'onoma de M\'exico, Ciudad de M\'exico, M\'exico }

\author{A.~Carrami\~{n}ana}
\affiliation{Instituto Nacional de Astrof\'isica, \'Optica y Electr\'onica, Puebla, M\'exico }

\author{F.~Carre\'on}
\affiliation{Instituto de Astronom\'ia, Universidad Nacional Aut\'onoma de M\'exico, Ciudad de M\'exico, M\'exico }

\author{S.~Casanova}
\affiliation{Instytut Fizyki Jadrowej im Henryka Niewodniczanskiego Polskiej Akademii Nauk, IFJ-PAN, Krakow, Poland }

\author{U.~Cotti}
\affiliation{Universidad Michoacana de San Nicolás de Hidalgo, Morelia, M\'exico }

\author{J.~Cotzomi}
\affiliation{Facultad de Ciencias F'isico Matemáticas, Benemérita Universidad Autónoma de Puebla, Puebla, M\'exico }

\author{S.~Couti\~{n}o de Le\'on}
\affiliation{Department of Physics, University of Wisconsin-Madison, Madison, WI, USA }

\author{E.~De la Fuente}
\affiliation{Departamento de F\'isica, Centro Universitario de Ciencias Exactas e Ingenier\'ias, Universidad de Guadalajara, Guadalajara, M\'exico }

\author{C.~de Le\'on}
\affiliation{Universidad Michoacana de San Nicol\'as de Hidalgo, Morelia, M\'exico }

\author{D.~Depaoli}
\affiliation{Max-Planck Institute for Nuclear Physics, 69117 Heidelberg, Germany}

\author{N.~Di Lalla}
\affiliation{Department of Physics, Stanford University: Stanford, CA 94305–4060, USA}

\author{R.~D\'iaz Hern\'andez}
\affiliation{Instituto Nacional de Astrof\'isica, \'Optica y Electr\'onica, Puebla, M\'exico }

\author{B.L.~Dingus}
\affiliation{Physics Division, Los Alamos National Laboratory, Los Alamos, NM, USA }

\author{M.A.~DuVernois}
\affiliation{Department of Physics, University of Wisconsin-Madison, Madison, WI, USA }

\author{K.~Engel}
\affiliation{Department of Physics, University of Maryland, College Park, MD, USA }

\author{T.~Ergin}
\affiliation{Department of Physics and Astronomy, Michigan State University, East Lansing, MI, USA }

\author{C.~Espinoza}
\affiliation{Instituto de F\'isica, Universidad Nacional Aut\'onoma de M\'exico, Ciudad de M\'exico, M\'exico }

\author{K.L.~Fan}
\affiliation{Department of Physics, University of Maryland, College Park, MD, USA }

\author{K.~Fang}
\affiliation{Department of Physics, University of Wisconsin-Madison, Madison, WI, USA }

\author{N.~Fraija}
\affiliation{Instituto de Astronom\'ia, Universidad Nacional Aut\'onoma de M\'exico, Ciudad de M\'exico, M\'exico }

\author{S.~Fraija}
\affiliation{Instituto de Astronom\'ia, Universidad Nacional Aut\'onoma de M\'exico, Ciudad de M\'exico, M\'exico }

\author{J.A.~Garc\'ia-González}
\affiliation{Tecnologico de Monterrey, Escuela de Ingenieria y Ciencias, Monterrey, N. L., M\'exico, 64849}

\author{F.~Garfias}
\affiliation{Instituto de Astronom\'ia, Universidad Nacional Aut\'onoma de M\'exico, Ciudad de M\'exico, M\'exico }

\author{H.~Goksu}
\affiliation{Max-Planck Institute for Nuclear Physics, 69117 Heidelberg, Germany}

\author{M.M.~Gonz\'alez}
\affiliation{Instituto de Astronom\'ia, Universidad Nacional Aut\'onoma de M\'exico, Ciudad de M\'exico, M\'exico }

\author{J.A.~Goodman}
\affiliation{Department of Physics, University of Maryland, College Park, MD, USA }

\author{S.~Groetsch}
\affiliation{Department of Physics, Michigan Technological University, Houghton, MI, USA }

\author{J.P.~Harding}
\affiliation{Physics Division, Los Alamos National Laboratory, Los Alamos, NM, USA }

\author{S.~Hern\'andez-Cadena}
\affiliation{Tsung-Dao Lee Institute \& School of Physics and Astronomy, Shanghai Jiao Tong University}

\author{I.~Herzog}
\affiliation{Department of Physics and Astronomy, Michigan State University, East Lansing, MI, USA }

\author{J.~Hinton}
\affiliation{Max-Planck Institute for Nuclear Physics, 69117 Heidelberg, Germany}

\author{D.~Huang}
\affiliation{Department of Physics, University of Maryland, College Park, MD, USA }

\author{F.~Hueyotl-Zahuantitla}
\affiliation{Universidad Autónoma de Chiapas, Tuxtla Gutiérrez, Chiapas, México}

\author{P.~H{\"u}ntemeyer}
\affiliation{Department of Physics, Michigan Technological University, Houghton, MI, USA }

\author{A.~Iriarte}
\affiliation{Instituto de Astronom\'{i}a, Universidad Nacional Aut\'onoma de M\'exico, Ciudad de M\'exico, M\'exico }

\author{S.~Kaufmann}
\affiliation{Universidad Polit\'ecnica de Pachuca, Pachuca, Hgo, M\'exico }

\author{A.~Lara}
\affiliation{Instituto de Geof\'isica, Universidad Nacional Autónoma de México, Ciudad de M\'exico, M\'exico }

\author{J.~Lee}
\affiliation{University of Seoul, Seoul, Republic of Korea}

\author{H.~Le\'on Vargas}
\affiliation{Instituto de F\'isica, Universidad Nacional Aut\'onoma de M\'exico, Ciudad de M\'exico, M\'exico }

\author{J.T.~Linnemann}
\affiliation{Department of Physics and Astronomy, Michigan State University, East Lansing, MI, USA }

\author{A.L.~Longinotti}
\affiliation{Instituto de Astronom\'ia, Universidad Nacional Aut\'onoma de M\'exico, Ciudad de M\'exico, M\'exico }

\author{G.~Luis-Raya}
\affiliation{Universidad Polit\'ecnica de Pachuca, Pachuca, Hgo, M\'exico }

\author{K.~Malone}
\affiliation{Physics Division, Los Alamos National Laboratory, Los Alamos, NM, USA }

\author{J.~Mart\'inez-Castro}
\affiliation{Centro de Investigaci\'on en Computaci\'on, Instituto Polit\'ecnico Nacional, M\'exico City, M\'exico.}

\author{J.A.~Matthews}
\affiliation{Dept of Physics and Astronomy, University of New M\'exico, Albuquerque, NM, USA }

\author{P.~Miranda-Romagnoli}
\affiliation{Universidad Autónoma del Estado de Hidalgo, Pachuca, M\'exico }

\author{J.A.~Montes}
\affiliation{Instituto de Astronom\'{i}a, Universidad Nacional Aut\'onoma de M\'exico, Ciudad de M\'exico, M\'exico }

\author{E.~Moreno}
\affiliation{Facultad de Ciencias F\'{i}sico Matem\'aticas, Benem\'erita Universidad Aut\'onoma de Puebla, Puebla, M\'exico }

\author{M.~Mostaf\'a}
\affiliation{Department of Physics, Temple University, Philadelphia, Pennsylvania, USA}

\author{L.~Nellen}
\affiliation{Instituto de Ciencias Nucleares, Universidad Nacional Autónoma de M\'exico, Ciudad de M\'exico, M\'exico }

\author{M.U.~Nisa}
\affiliation{Department of Physics and Astronomy, Michigan State University, East Lansing, MI, USA }

\author{R.~Noriega-Papaqui}
\affiliation{Universidad Aut\'onoma del Estado de Hidalgo, Pachuca, M\'exico }

\author{L.~Olivera-Nieto}
\affiliation{Max-Planck Institute for Nuclear Physics, 69117 Heidelberg, Germany}

\author{N.~Omodei}
\affiliation{Department of Physics, Stanford University: Stanford, CA 94305–4060, USA}

\author{M. Osorio}
\affiliation{Instituto de Astronom\'{i}a, Universidad Nacional Aut\'onoma de M\'exico, Ciudad de M\'exico, M\'exico }

\author{Y.~P\'erez Araujo}
\affiliation{Instituto de F\'{i}sica, Universidad Nacional Aut\'onoma de M\'exico, Ciudad de M\'exico, M\'exico }

\author{E.G.~P\'erez-P\'erez}
\affiliation{Universidad Polit\'ecnica de Pachuca, Pachuca, Hgo, M\'exico }

\author{C.D.~Rho}
\affiliation{Department of Physics, Sungkyunkwan University, Suwon 16419, South Korea}

\author{D.~Rosa-Gonz\'alez}
\affiliation{Instituto Nacional de Astrof\'isica, \'Optica y Electr\'onica, Puebla, M\'exico }

\author{E.~Ruiz-Velasco}
\affiliation{Max-Planck Institute for Nuclear Physics, 69117 Heidelberg, Germany}

\author{H.~Salazar}
\affiliation{Facultad de Ciencias F\'{i}sico Matem\'aticas, Benem\'erita Universidad Aut\'onoma de Puebla, Puebla, M\'exico }

\author{D.~Salazar-Gallegos}
\affiliation{Department of Physics and Astronomy, Michigan State University, East Lansing, MI, USA }

\author{A.~Sandoval}
\affiliation{Instituto de F\'isica, Universidad Nacional Aut\'onoma de M\'exico, Ciudad de M\'exico, M\'exico }

\author{M.~Schneider}
\affiliation{Department of Physics, University of Maryland, College Park, MD, USA }

\author{G.~Schwefer}
\affiliation{Max-Planck Institute for Nuclear Physics, 69117 Heidelberg, Germany}

\author{J.~Serna-Franco}
\affiliation{Instituto de F\'isica, Universidad Nacional Aut\'onoma de M\'exico, Ciudad de M\'exico, M\'exico }

\author{A.J.~Smith}
\affiliation{Department of Physics, University of Maryland, College Park, MD, USA }

\author{Y.~Son}
\affiliation{University of Seoul, Seoul, Republic of Korea}

\author{R.W.~Springer}
\affiliation{Department of Physics and Astronomy, University of Utah, Salt Lake City, UT, USA }

\author{O.~Tibolla}
\affiliation{Universidad Polit\'ecnica de Pachuca, Pachuca, Hgo, M\'exico }

\author{K.~Tollefson}
\affiliation{Department of Physics and Astronomy, Michigan State University, East Lansing, MI, USA }

\author{I.~Torres}
\affiliation{Instituto Nacional de Astrof\'isica, \'Optica y Electr\'onica, Puebla, M\'exico }

\author{R.~Torres-Escobedo}
\affiliation{Tsung-Dao Lee Institute \& School of Physics and Astronomy, Shanghai Jiao Tong University}

\author{R.~Turner}
\affiliation{Department of Physics, Michigan Technological University, Houghton, MI, USA }

\author{F.~Ure\~{n}a-Mena}
\affiliation{Instituto Nacional de Astrof\'isica, \'Optica y Electr\'onica, Puebla, M\'exico }

\author{E.~Varela}
\affiliation{Facultad de Ciencias F\'{i}sico Matem\'aticas, Benem\'erita Universidad Aut\'onoma de Puebla, Puebla, M\'exico }

\author{X.~Wang}
\affiliation{Department of Physics, Michigan Technological University, Houghton, MI, USA }

\author{I.J.~Watson}
\affiliation{University of Seoul, Seoul, Republic of Korea}

\author{K.~Whitaker}
\affiliation{Department of Physics, Pennsylvania State University, University Park, PA, USA }

\author{E.~Willox}
\affiliation{Department of Physics, University of Maryland, College Park, MD, USA }

\author{H.~Wu}
\affiliation{Department of Physics, University of Wisconsin-Madison, Madison, WI, USA }

\author{S.~Yu}
\affiliation{Department of Physics, Pennsylvania State University, University Park, PA, USA }

\author[   0000-0002-9307-0133]{S.~Yun-C\'arcamo}
\affiliation{Department of Physics, University of Maryland, College Park, MD, USA }

\author{H.~Zhou}
\affiliation{Tsung-Dao Lee Institute \& School of Physics and Astronomy, Shanghai Jiao Tong University}

\collaboration{96}{HAWC Collaboration}

%\author{Butler Burton}
%\affiliation{Leiden University}
%\affiliation{AAS Journals Associate Editor-in-Chief}

%\altaffiliation{AASTeX v6+ programmer}

%\affiliation{AAS Director of Publishing}
%\affiliation{American Astronomical Society \\
%1667 K Street NW, Suite 800 \\

%\author{Magaret Donnelly}
%\affiliation{IOP Publishing, Washington, DC 20005}



\begin{abstract}

The High-Altitude Water Cherenkov (HAWC) Gamma-Ray Observatory located on the side of the Sierra Negra volcano in Mexico, has been fully operational since 2015. The HAWC collaboration has recently significantly improved their extensive-air-shower reconstruction algorithms, which has notably advanced the observatory performance. The energy resolution for primary gamma rays with energies below 1~TeV was improved by including a noise-suppression algorithm. Corrections have also been made to systematic errors in direction fitting related to the detector and shower plane inclinations, $\mathcal{O}(0.1^{\circ})$ biases in highly inclined showers, as well as enhancements to the core reconstruction. The angular resolution for gamma rays approaching the HAWC array from large zenith angles ($> 37^{\circ}$) has improved by a factor of four at the highest energies ($> 70$~TeV) as compared to previous reconstructions. The inclusion of a lateral distribution function fit to the extensive air shower footprint on the array to separate gamma-ray primaries from cosmic-ray ones, based on the resulting $\chi^{2}$ values, improved the background rejection performance at all inclinations. At large zenith angles, the improvement in significance is a factor of four compared to previous HAWC publications. These enhancements have been verified by observing the Crab Nebula, which is an overhead source for the HAWC Observatory. We show that the sensitivity to Crab-like point sources ($E^{-2.63}$) with locations overhead to 30$^{\circ}$ zenith is comparable or less than 10\% of the Crab Nebula's flux between 2 and 50~TeV. Thanks to these improvements, HAWC can now detect more sources, including the Galactic Center.

\end{abstract}

%\keywords{ keywords}


\section{Introduction} \label{sec:intro} 

Astrophysical gamma rays play a crucial role in multi-messenger astronomy. Detecting Very-High-Energy and Ultra-High-Energy (VHE/UHE) gamma rays allows the study of particle acceleration processes beyond the energy reach of Earth-bound accelerators. Unlike charged cosmic rays accelerated by astrophysical sources, gamma rays produced near these sites are not deflected by interstellar magnetic fields and thus can be traced back to the accelerator. Gamma rays in the VHE/UHE regime also complement the information provided by the detection of gravitational waves and neutrinos \citep{hinton2020multi}. They uniquely contribute to the investigation of explosive transients and the search for Dark Matter or Lorentz Invariance Violation at otherwise inaccessible energy and distance scales.

Direct detection of gamma rays can only occur through space-based detectors. While these are well-suited for gamma-ray energies up to hundreds of GeV, cost and feasibility limit the effective detector area in space and thus their capability of higher energy detection. Ground-based detectors indirectly observe gamma rays by detecting the secondary particles of Extensive Air Showers (EASs) \citep{matthews2005heitler}, initiated when the gamma rays interact with nuclei in Earth's atmosphere. One EAS detection technique is via Imaging
Air Cherenkov Telescopes (IACTs), which observe the Cherenkov radiation that is generated by the secondary charged particles of the EAS traveling faster than the speed of light in the Earth's atmosphere. However, their duty cycle ($\sim 15$\%) and field of view (e.g. 5$^{\circ}\times 5^{\circ}$) are small  \citep{de2018gamma,abdalla2021tev}. The High-Altitude Water Cherenkov (HAWC) Observatory is based on the water-Cherenkov technique, where photomultiplier tubes (PMTs) placed in water tanks detect Cherenkov radiation generated by the secondary charged particles in the water, rather than in the atmosphere. To obtain the estimated energy and direction of the primary gamma ray, the EAS of secondary particles is reconstructed employing only the information measured from these Cherenkov-radiation photons: measured photoelectrons, arrival time, and spatial distribution. However, over 99.9\% of the EASs detected are not produced by gamma rays, but hadrons. This fact raises one of the challenges faced during the event-reconstruction stage: distinguishing between the cosmic-ray background and gamma-ray-produced showers \citep{bose2021ground}. Hence, the performance of the HAWC Observatory is thoroughly shaped by the methods and algorithms employed when reconstructing EAS events. 

The HAWC Observatory consists of two arrays of water Cherenkov detector (WCD) stations: a primary array of 300 densely packed large-volume WCDs located at the center (``primary detector'') surrounded by a sparse outer array of 345 small-volume WCDs (``outriggers'') \citep{abeysekara2022characterizing}. This paper presents the performance of the primary detector and associated data reconstruction and analysis. The primary detector has been operational
since 2015 on the flanks of the Sierra Negra volcano in Mexico. Successfully detecting and reconstructing gamma-ray events depends on several conditions and parameters to be satisﬁed, mainly: PMT trigger conditions and quality assessment, the
reconstruction precision and accuracy of the shower core location (i.e. the intersection point of the primary particle direction with the array plane), the primary particle direction and energy, and the effectiveness of the gamma/hadron separation. All results published by the HAWC Collaboration since 2017, including the 2HWC \citep{abeysekara20172hwc} and 3HWC catalogs \citep{albert20203hwc}, were based on the fourth version (``Pass 4") of the HAWC event-reconstruction algorithms, which has been extensively described in previous publications \citep{smith2015hawc,abeysekara2017observation,abeysekara20172hwc,abeysekara2018data,abeysekara2019measurement}. 

Here we present a substantial update of the algorithms used to reconstruct data from the primary detector of the HAWC instrument, referred to as the ``Pass 5" version, which was completed in 2023 and will be used for upcoming papers. The main improvements are the extension of analyses to lower energies with new methodologies to identify and remove the background, the enhancement of the directional reconstruction for high-energy showers, and the improvement of the gamma/hadron separation efficiency, especially for showers approaching the array from high zenith angles (angular difference of arrival directions from zenith). The paper is organized as follows. A short description of the HAWC detector and its simulation in Section \ref{sec:hawc} is followed by a discussion of the algorithms used for Pass 5 event reconstruction in Section \ref{sec:rec} and changes in the gamma/hadron separation cut optimization. Section \ref{sec:perf} presents comparisons of Pass 4 and Pass 5 reconstruction performance using simulated data. The
verification of Pass 5 performance with data from the Crab nebula as a reference source is reported
in Section \ref{sec:crab}, followed by a study of systematic uncertainties in Section \ref{sec:syst}. The paper closes with a description of the differential sensitivity achieved by the Pass 5 reconstruction in Section \ref{section:sensitivity} and conclusions in Section \ref{sec:conc}.

\section{The HAWC detector array and simulation} \label{sec:hawc}
The HAWC Observatory is located at an altitude of 4100 m above sea level and 18$^{\circ}$59.7$^\prime$ North, 97$^{\circ}$19.0$^\prime$ West. The primary array was constructed between 2011 and 2014 and started to fully operate in March 2015. At any given instant, HAWC observes 15\% of the overhead sky (2 sr), covering two-thirds of the sky in every 24-hour cycle \citep{abeysekara2018data}. The primary array consists of 300 cylindrical steel tanks---5.4 meters high and 7.3 meters in diameter---that contain large PVC bladders, deployed over a total physical area of approximately 22,000 m$^2$. Each bladder contains four upward-facing PMTs: one 10$"$ high-quantum efficiency Hamamatsu R7081 PMT at the center and three 8$"$ Hamamatsu R5912 PMTs at the corners of a centered 3.2 meter-side equilateral triangle. The bladders are filled to 4.5 m depth of purified water, with 4 m above the PMTs \citep{ABEYSEKARA2023168253}. The distribution of the tanks is a compromise between maximizing the density for shower particle detection and serviceability.

The HAWC reconstruction and analysis algorithms are optimized for primary gamma rays with energies from $\sim$~300~GeV to several hundred TeV. The HAWC simulation includes different stages, from generating the air shower to the PMTs response. Air shower events are simulated with CORSIKA (COsmic Ray SImulations for KAscade \citep{heck1998corsika}). Simulated primary particles include gamma rays and cosmic-ray protons, as well as  helium, carbon, iron, magnesium, neon, oxygen, and silicon nuclei. Events are simulated with energies from 5 to 5$\times 10^5$~GeV and thrown in higher quantities at small distances from the detector center (up to a 1 km radius) than elsewhere. CORSIKA output files contain information about the secondary particles at the altitude of the HAWC Observatory. The passage of these particles through the WCDs until they produce photoelectrons (PEs) in the PMTs is simulated specifically for HAWC hardware using GEANT4 \citep{allison2007facilities}. After the event reconstruction---as will be described in the following section---events are weighted based on their species and core distance. Additionally, to increase the statistics of simulated high-energy events, we weight events on a hard spectrum by default but can re-weight them to simulate the detector response to any spectrum. More details about the primary detector hardware and simulation can be found in \citet{ABEYSEKARA2023168253}.

\section{Extensive-Air-Shower Event Reconstruction} \label{sec:rec}

We reconstruct the angle of the symmetry axis of the incoming shower (which represents the direction of the primary particle), the primary particle energy, and the lateral distribution function of the shower to distinguish gamma-ray-induced from hadron-induced showers. This is accomplished using times when the EAS front strikes the PMTs and the amount of charge detected, which is measured in recorded PEs. In order to reconstruct these air shower features, it is crucial to identify two distinguishing characteristics: the shower core and plane. The former is the projection of the densest region of the shower front onto the detector plane around the intersection point with the shower axis. For the latter feature, although the shower front is curved, it is convenient to identify a main shower plane that accounts for the inclination of the front of secondary particles hitting the PMTs. To account for this curvature in the direction reconstruction, it is critical to estimate accurately the core location. The shower direction is determined by fitting the shape of the shower front with respect to the core location \citep{abeysekara2019measurement}. We have two independent energy estimators that are used interchangeably: the ground parameter, which fits the lateral distribution of the showers, and a neural network method, which mainly uses the charge distribution in annuli around the shower core as input. More information can be found in \citep{abeysekara2019measurement}. Lastly, the gamma/hadron separation algorithms generally exploit the differences in how electromagnetic energy and secondary muons and hadrons are distributed with respect to the core location. The showers induced by gamma-ray primaries are usually purely electromagnetic and lead to footprints in the array that have ``smoother'' charge signal distributions than hadron-induced showers. Array footprints of hadron showers are characterized by more variable charge signals/clusters, especially farther away from the shower core.

In this section, we present the improvements in after-pulse veto at high energies, noise suppression, core location identification, angle and energy reconstruction, and gamma/hadron separation in the Pass 5 analysis. We provide a summary of the HAWC reconstruction steps in Table \ref{tab:reco}. 

\begin{table}[ht!]
\begin{center}
\begin{tabular}{ |c|c|} 
 \hline
Step & Description \\
 \hline
1. & Event diagnostic and filter \\
2.& Hit selection by noise-suppression algorithm\\
3.& Center-of-Mass core estimation and direction reconstruction \\
4.& Core reconstruction\\
5.& Direction reconstruction \\
6.& Energy reconstruction \\
7.& Reduction of data\\
8.& Corrections to direction reconstruction  \\
9.& Gamma/hadron separation by bins\\
 \hline
 \end{tabular}
\caption{Summary of event reconstruction in HAWC: {\textbf{1.}} We first filter events with short-term variability in the PE
distribution of the PMTs (e.g. due to lightning) and after-pulsing effects (discussed in Section \ref{subsec:apulse}). {\textbf{2.}} We remove noise hits as described in Section \ref{subsec:lowe}. {\textbf{3.}} We make a preliminary estimate of the center-of-mass of the charge detected by the PMTs \citep{abeysekara2017observation} and initial direction to use as input to the next steps. {\textbf{4.}} We perform the final core reconstruction as described in Section \ref{subsec:core}. {\textbf{5.}} We conduct the final direction reconstruction as described in Section \ref{subsec:dir}. {\textbf{6.}} We carry out the energy reconstruction as described in Section \ref{subsec:energy}. {\textbf{7.}} We reduce the data with a loose gamma/hadron separation cut and group time periods where the detector was stable. {\textbf{8.}} We add the corrections described in Section \ref{subsec:dir}. {\textbf{9.}} We bin the data and apply gamma/hadron separation cuts as described in Section \ref{subsec:ghcuts} and \ref{subsec:cutopt}.\label{tab:reco} }
\end{center}
\end{table}

\subsection{High-energy after-pulse veto}\label{subsec:apulse}

After-pulses caused by ionized residual gas molecules inside the PMT are occasionally observed in HAWC PMTs within 10 to 15 microseconds after a hit \citep{abeysekara2022characterizing}. After a large hit, we veto PMTs with after-pulses arriving during this time window by treating the PMTs as unavailable for the following reconstruction steps. In Pass 4, we required availability of at least 90\% of the PMTs operating to carry out the reconstruction. We found this to be too restrictive for high-energy events where saturation causes PMTs be unavailable 
 and it resulted in a significant data loss. In Pass 5, we required only 80\% of the PMTs be available, allowing us to recover the data without compromising the performance of the reconstruction, as shown later in Section \ref{sec:perf}.

\subsection{Noise suppression}\label{subsec:lowe}

Air-shower events in HAWC are triggered and logged to disk when at least 28 of the 1200 PMTs are hit within a 150~ns window. This low threshold was set before the event rate starts to rapidly increase as a function of the number of PMTs hit to reduce the stored raw data stream \citep{ABEYSEKARA2023168253}. However, near this lower limit noise events from randomly coincident air showers can also trigger the detector. When a trigger is identified, PMT hits within a 1500~ns window (500~ns before the trigger and 1000~ns after) are recorded. We consider the first hit in each channel as valid, so a PMT can only have a single hit per event. In previously published results based on the Pass 4 event reconstruction, we utilized only showers with more than 80 hits (the Pass 4 Bin 1 threshold, see Table \ref{tab:bins}). This excluded about 80\% of our triggers since these noise hits tended to compromise the direction fits and the gamma/hadron separation algorithms. At lower energies, they limited the reconstruction quality of small events. At higher energies, they contributed to the noise through the misidentification of multiple small showers as one big event.

In Pass 5, we introduced a cleaning algorithm before the main reconstruction stage called the Multi-Plane Fitter (MPF) \citep{RosThesis}. The MPF assigns hits to different shower planes based on their arrival time at the PMTs. The algorithm adds an additional shower plane if its addition improves the likelihood using a Bayesian information criterion (BIC)\citep{schwarz1978estimating},
\begin{equation}\label{eq:mpf}
    \Delta \text{BIC}=-2\Delta L+k(\log(n)+\log(2\pi))>0\,,
\end{equation}
where $\Delta L$ is the log-likelihood difference, $k$ is the dimensionality difference between models (in this case $k=3$), and $n$ is the number of hits. Finally, the MPF treats all hits associated with the most significant shower plane found as the air-shower signal in that event window and discards all other hits on any additional planes as noise. Applying this algorithm improves the reconstruction quality enough to warrant cleaned small events inclusion in the analysis (see Table \ref{tab:bins}). This greatly improves the sensitivity to gamma-ray bursts, active galaxy nuclei, and galactic gamma-ray sources emitting fluxes at the detection threshold \citep{Yuan_2022,galaxies10020039}.

\begin{table}[ht!]
\begin{center}
\begin{tabular}{ |c|c|c|c|c|c|} 
 \hline
Pass 4 & Hit PMT fraction & Pass 5 & Hit PMT fraction & Median energy Crab  &Median energy Crab \\
& & & &on array (TeV) & off array (TeV) \\
 \hline
 ---& --- & B0  & 2.7  -- 4.7 \% &0.28 &0.57\\
---& ---&B1  & 4.7  -- 6.8 \% & 0.38& 0.88\\
B1&  6.7 -- 10.5\% &B2  & 6.8  -- 10.4\% & 0.53&1.29\\
B2&10.5 -- 16.2\% &B3  & 10.4 -- 16.1\% & 0.83&2.02\\
B3&16.2 -- 24.7\%  &B4  & 16.1 -- 24.5\% &1.37 &3.66\\
B4& 24.7 -- 35.6\% &B5  & 24.5 -- 35.1\% &2.25 &6.21\\
B5& 35.6 -- 48.5\% &B6  & 35.1 -- 47.2\% & 3.68&10.27\\
B6& 48.5 -- 61.8\% &B7  & 47.2 -- 59.9\% & 5.97&16.62\\
B7&  61.8 -- 74.0\% &B8  & 59.9 -- 72.2\% &9.54 &25.78\\
B8&74.0 -- 84.0\%  &B9  & 72.2 -- 82.2\% &14.63 &41.47\\
B9& 84.0 -- 100.0\% &B10 & 82.2 -- 100.0\% &30.46 &73.91\\
 \hline
 \end{tabular}
\caption{Comparison of the HAWC Pass 5 data bins with the previous Pass 4 definitions. As presented in Section 4.2 of \citet{abeysekara2019measurement}, the main binning of HAWC data is according to the fraction of PMTs hit (FHit). The edges of the bins were chosen using data such that from one bin to the following the number air shower events (both hadronic and gamma-ray induced) reduces approximately by half. As a reference, we include the median of the energy distribution observed from the Crab Nebula, for each Pass 5 FHit bin. Thanks to the introduction of the MPF in the Pass 5 version, we added two bins at the low-energy end of the HAWC gamma-ray detection range, significantly increasing the sensitivity of HAWC analyses below 1 TeV.}\label{tab:bins} 
\end{center}
\end{table}

\subsection{Core reconstruction}\label{subsec:core}

In the Pass 4 analyses, we fit the core location by performing an iterative numerical approximation using Newton's method to fit the charges of hit PMTs to a lateral distribution function (LDF) consisting of a Gaussian of fixed $\sigma=10$~m close to the core location combined with a tail further away from it that falls like $1/r^3$ \citep{abeysekara2017observation}. This method is fast compared to a fit utilizing the Nishimura-Kamata-Greisen (NKG) function \citep{kamata1958lateral} and simulations show that it accurately identifies the location of the core for events whose core lands on the primary array. With Pass 5, we use a much more sophisticated algorithm than the simple $\chi^2$ approach used in Pass 4. In the new approach, we use simulated gamma-ray showers to generate probability density functions (PDFs) for observing hits at a given amplitude (in PEs), radius from the core, angle between the zenith and the reconstructed shower direction, and true primary particle energy. We then use the PDFs to perform a maximum likelihood fit to identify the core location. The inherent uncertainty in the initial interaction of the primary gamma-ray brings some ambiguity to determining the core location since different combinations of the PDF parameters may reproduce the same shower characteristics. Including PMTs that were available but unhit (measured 0 PEs) in the PDFs gives a much more reliable model for the observed PE fluctuations, helping to reduce the ambiguity by maximizing the amount of data for the fit \citep{Joshi_2019}.

In addition to the implementation of the PDFs, the Pass 5 core fitter recognizes whether the shower core fell on or off the primary array. Pass 4 FHit analyses considered all events regardless of core location. In Pass 5 we define ``off-array'' events with a core location between the outer edge of the detector and a concentric area equal to 1.5 times the area of the main array. While off-array events are reconstructed with poorer angular resolution than on-array events, they contribute to HAWC sensitivity. On-array and off-array events are divided into different bins that share the same FHit boundaries (see Table \ref{tab:bins}) but have different gamma/hadron cuts. Note that in Pass 4 the energy estimator-binned analyses only considered on-array events and that continues to be true in Pass 5.
   
\subsection{Direction reconstruction}\label{subsec:dir}

The front of an electromagnetic EAS comprises mainly gamma rays, electrons, and positrons scattered away from the axis with an angle that depends on their momentum and the thickness of the air traversed. Hence, at a given altitude, the front of the developing shower is curved (convex), has a finite thickness and varying particle density, with the densest region near the axis. It is crucial to account for these features since the direction reconstruction relies on the information recorded from secondary particle hits: arrival time and charge in PEs. For a shower hitting the detector directly from above, the densest part of the shower (the core) will strike the PMTs first---due to its curvature---recording a large number of PEs. For a highly inclined shower, the first secondary particles to arrive correspond to the least dense part of the shower front, which may not cross the PMTs recording threshold until a higher number of PEs is detected causing a delay in the recording. We approximate the timing PDFs---for each amplitude measured in PEs---and the deviation of the shower front plane with second-order polynomials of the radial distance of the secondary particles to the shower core. The curvature approximation and timing distributions are then used to fit the incoming shower direction.

While the shower-fitting methodology is very similar to Pass 4, in Pass 5, we apply after-fit corrections to address several small but salient systematic errors present in Pass 4. We identified a small error in the detector survey which did not properly account for the angle of the primary array plane with respect to the horizontal. The calibration procedure applies a minor adjustment to the absolute timing of each hit to guarantee that the zenith is the direction of the highest cosmic-ray shower rate, so we only see the tilt effect at large zenith angles. A second error was identified in the reconstructed zenith angle in both simulated and real data from the Crab. The effect introduces an offset that increases rapidly from below 0.1$^\circ$ to values greater than or equal to about 0.1--0.2$^\circ$ with increasing zenith angle. The exact cause is unknown, but it is likely due to incorrectly simulating particles when they pass through the sides of the WCDs. These two errors are corrected by a zenith-angle-dependent shift for all shower directions. As with the shower core reconstruction, the shower curvature correction is also now applied in the shower plane instead of the detector plane (which in Pass 4 introduced a bias in the reconstruction of inclined showers). The offset from this bias significantly degraded the angular resolution of extreme declination sources culminating at large zenith angles and was found to depend on the shower core projection (the distance between the core location and a line perpendicular to the shower direction passing through the center of the array). These corrections drastically improved the direction reconstruction for high zenith angle showers, as will be discussed in Section \ref{subsubsection:angle}.

\subsection{Energy reconstruction}\label{subsec:energy}

In Section 3 of \citet{abeysekara2019measurement}, we described two energy estimator algorithms that we currently use in the analysis of HAWC data to estimate the energy of the primary gamma ray from the charge density and shower age. The Ground Parameter (GP) performs a fit to the lateral distribution to measure the charge density at 40~m from the core location for on-array events (65 m if off-array) and uses the charge density along with the zenith angle of the event to estimate the energy. The Neural Network (NN) uses, among other things, the fractions of charge deposited in nine concentric annuli around the shower core location---every 10 m in radius---as inputs. As in Pass 4 and described in Section 3 of \citet{abeysekara2019measurement}, we divide the on-array data sample depending on the estimated energy of the primary gamma rays into the same twelve analysis bins for both GP and NN in the range from $10^{2.5}$ to $10^{5.5}$~GeV (see Table \ref{tab:enbins}).

\begin{table}[ht!]
\begin{center}
\begin{tabular}{ |c|c|c|} 
 \hline
 Energy bins  & Min. $\log_{10}(E/\text{GeV})$  & Max. $\log_{10}(E/\text{GeV})$ \\
 \hline
 Ea & 2.50 & 2.75 \\
 Eb  & 2.75 & 3.00  \\
 Ec & 3.00 & 3.25 \\
 Ed  & 3.25  & 3.50  \\
 Ee & 3.50 & 3.75 \\
 Ef  &  3.75 & 4.00  \\
 Eg & 4.00 & 4.25 \\
 Eh  & 4.25  & 4.50  \\
 Ei & 4.50 & 4.75 \\
 Ej  & 4.75  & 5.00  \\
 Ek & 5.00 & 5.25 \\
 El & 5.25  & 5.50\\
 \hline
 \end{tabular}
\caption{HAWC energy bin definitions for GP and NN as described in \citep{abeysekara2019measurement}. See Section \ref{subsec:energy} for more details about energy estimation.\label{tab:enbins} }
\end{center}
\end{table}

\subsection{Gamma/hadron separation}\label{subsec:ghcuts}

Gamma rays and hadrons produce EASs when they hit the upper atmosphere. However, their development differs in secondary particle species and shower shape. Hadron-induced showers contain many muons with high transverse momenta with respect to the shower axis, resulting in considerable energy deposited far from the core. In contrast, gamma-ray showers produce electromagnetic cascades with low-energy gamma rays and electrons populating the lateral tails. Consequently, the footprints of the EASs on the array are different depending on the shower progenitor--- hadronic showers produce irregular, asymmetric footprints while gamma-ray-induced shower footprints are axially symmetric. There are also showers initiated by cosmic-ray electrons, which we cannot differentiate from gamma-ray showers and constitute an irreducible (but small) background. Across the HAWC energy range, an average fraction of 6\% of the upper limit of HAWC diffuse gamma-ray background corresponds to cosmic-ray electron events \citep{hessecr,hawccollaboration2022limits}. 

Several algorithms have been developed to select gamma-ray showers over cosmic-ray ones based on the information measured on the ground. In Pass 4, we used two gamma/hadron separation parameters: compactness \citep{abeysekara2017observation} and PINCness (Parameter for Identifying Nuclear Cosmic rays). Compactness, which is retained by Pass 5, is calculated as the number of PMTs hit in an EAS event divided by the maximum charge measured by a PMT outside a radius of 40 m around the core location. A small value of compactness is an indication of the occurrence of a muon with high transverse momentum in an EAS event and hence of a hadronic primary. The latter parameter, PINC---a $\chi^{2}$-like measure of the smoothness of the charge footprint of a shower---is replaced in Pass 5 with the reduced $\chi^{2}$ obtained from the modified LDF fit performed in the determination of the ground parameter energy estimate in the primary array plane. Because the NKG function was originally conceived to describe purely electromagnetic EASs initiated by gamma-ray primaries, smaller $\chi^{2}$ 
 values tend to be associated with gamma rays, while large values indicate a higher probability for the shower to be produced by a hadron primary \citep{krawczynski2006gamma,alfaro2022gamma}.

\subsection{Cut optimization}\label{subsec:cutopt}

In practice, gamma/hadron separation is a filter we apply to the data. The cut values are selected such that they optimize the background rejection and gamma-ray retention in FHit and energy binning. This cut-optimization process was also improved in Pass 5 and here we describe the differences between the two analyses (summary in Table \ref{tab:ghopt}).

In Pass 4, there was no distinction in the core position for FHit binning. Hence, the optimization of gamma/hadron separation cuts was performed independently of the position of the core. In Pass 5, the cuts are optimized separately for on- and off-array events. Energy estimators are only optimized for on-array events in both Pass 4 and 5.

The FHit gamma/hadron cut optimization was performed using only data in Pass 4, by maximizing the Poisson significance of events near the Crab ($\pm5^{\circ}$). However, the signal-to-background ratio is small for lower FHit bins, making optimization noisy. In Pass 5, we simulate signal events weighted with a Crab-like spectrum ($E^{-2.63}$) and a flux norm of 3.45$\times 10^{-11}$~TeV/cm$^{2}$s at 1 TeV. The optimization consists of maximizing the signal-to-background ratio among different cut combinations. This is done for FHit bins B0 to B8 on-array and B0 to B6 off-array bins. For bins B9 to B10 on-array and B7 to B10, off-array we continue using data. The compactness parameter is only used in FHit bins that are optimized using simulations. Energy estimators are also optimized using simulations, but bins B7 (all energies) to B10 (except the three highest energy bins) use the same cuts as FHit on-array bins, where the compactness parameter is removed. For the three highest energy bins of B10 the compactness parameter is included to maintain a similar background level to other bins.

There are two sets of cuts for FHit bins: one set optimized for bright flaring (Crab-like) sources and one optimized for weak sources. For the latter, we optimized the expected significance of
 source with 2\% of the flux of the Crab, which makes the cuts tighter. Energy estimator cuts are optimized using simulations weighted with a full Crab flux, as in Pass 4.

The reduced LDF-$\chi^{2}$ threshold optimized on the Crab---an overhead source for HAWC---was too strict for other zenith angle sources at high energies (high FHit), causing a significant loss of effective area and fraction of gamma-ray induced events (see Section \ref{subsubsection:area} and \ref{subsubsection:gheff}). To address this, we included zenith-dependent cuts in some bins:
\begin{equation}\label{zenith}
    C(\theta)= C_{0}(1+A\theta^{4})\,,
\end{equation}
where the parameters $C_{0}$ and $A$ are optimized to determine the reduced LDF-$\chi^{2}$ threshold (cut) for each bin from B7 to B10, in all analysis approaches. Equation \ref{zenith} was found by fixing the efficiency of the cut for each bin as a function of $\theta$.

When the optimization is performed, we set a lower limit for the gamma-ray retention. An 80\% gamma-ray efficiency is required for NN and GP analyses to minimize the systematic errors that arise in these analyses (see Section \ref{sec:syst}). We set a looser lower limit for FHit (50\%) to account for sources that we can see near detection threshold but that may present larger systematic errors during the analyses.

\begin{table}[ht!]
\begin{center}
\begin{tabular}{ |c|c|c| } 
 \hline
Optimization & Pass 4  & Pass 5\\
 \hline
On-array core& FHit (no cut in position), NN, GP & FHit, NN, and GP   \\
Off-array core & FHit (no cut in position) & FHit  \\
\hline
Data & FHit & B9--10 all on-array; B7--10 off-array  \\
Simulation & NN and GP & All other bins \\
\hline
Compactness & all bins & NN and GP, FHit only in bins optimized\\
& & with simulations\\
PINCness & all bins & no  \\
LDFChi2 & no &  all bins  \\
\hline
100\% Crab& FHit, NN, GP & all FHit and NN/GP  \\
2\% Crab & FHit & all FHit
 \\
\hline
Zenith & overhead optimization & zenith dep. LDF-$\chi^{2}$ B7--10 all  \\
\hline
Gamma eff.& no restriction & 80\% NN \& GP, 50\% FHit \\
 \hline
 \end{tabular}
 \caption{Comparison of cut optimization between the Pass 4 and Pass 5 versions. See Section \ref{subsec:cutopt} for details about cut optimization, Table \ref{tab:bins} for FHit bin definitions, and Table \ref{tab:enbins} for energy bin definitions. \label{tab:ghopt}}
 \end{center}
 \end{table}

\section{Pass 5 performance tests using simulations}\label{sec:perf}

The following subsections show the
performance of the Pass 5 reconstruction algorithms compared for events reconstructed with cores landing on the primary array only, as was used in Pass 4. In addition, we show the performance of the reconstruction algorithms for events landing off-array which are newly added in Pass 5 (described in Section \ref{subsec:core}). We obtain results for the effective area, angular resolution, and hadron efficiency. We find gamma-efficiency from Monte Carlo (MC) simulation events. The performance is shown for three different zenith intervals, equally spaced in $\cos(\text{zenith angle})$ with limits 1.0, 0.9, 0.8, and 0.7, so that events approximately come from equal solid angle subtended in each zenith-angle interval with limits $0^{\circ}$, $26^{\circ}$, $37^{\circ}$, and $46^{\circ}$. 

\subsection{Effective area}\label{subsubsection:area}
%%%%%%%%%%%%%%%%%%%%%%%%%%%%%%%%%%%%%%%%%%%%%%%%%%%%%%%%%%%%%5

\begin{figure}%[ht!]
\centering
% \begin{center}
\resizebox{1.0\textwidth}{!}{%
\begin{tabular}{cc}
\includegraphics[width=0.5\textwidth]{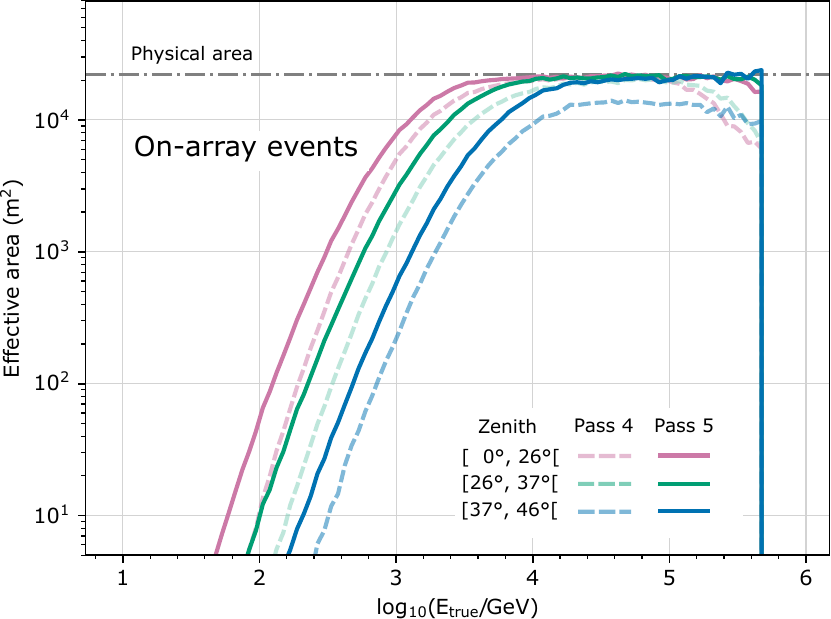}%
&\quad
\includegraphics[width=0.5\textwidth]{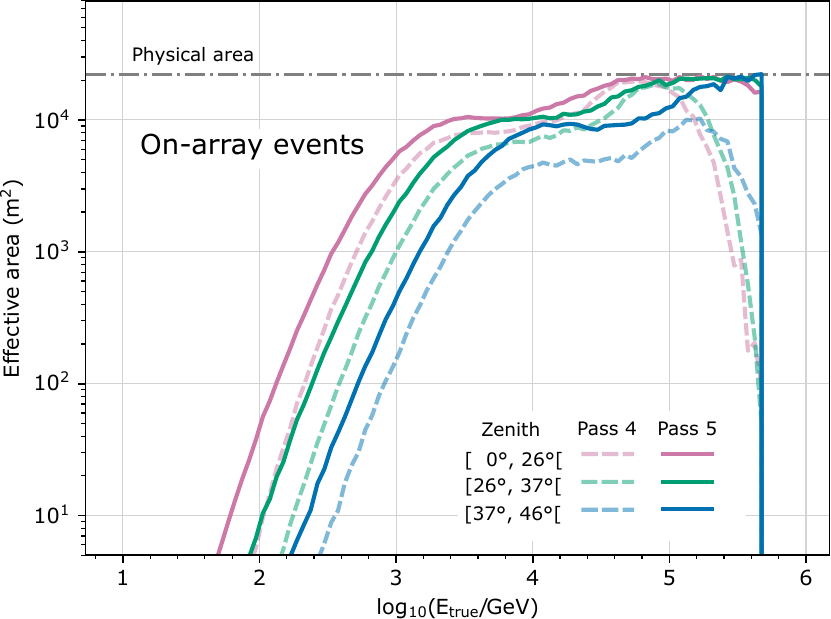}%
\\
(a) Trigger conditions & (b) Gamma/hadron cuts\\
\end{tabular}
}
\caption{Comparison between the FHit effective area of the Pass 4 and Pass 5 EAS reconstructions for three different zenith-angle intervals for \textbf{(a)} events in which the shower core lands on the array and which pass the trigger conditions and \textbf{(b)} events that additionally pass gamma/hadron cuts. We stop plotting the effective area at 500 TeV because at higher energy the PMTs receive charge beyond their calibration limit (hardware saturation).  With the Pass 5 improvements, useable effective area begins at a lower energy than in Pass 4, and at the highest energy, the effective area saturates at the physical area of the primary array. The Pass 4 performance is not the full Pass 4 performance, only of on-array Pass 4 events.\label{fig:areaon}}
% \end{center}
\end{figure}

We calculate the effective collection area of the primary HAWC array, as a function of energy, using simulations (see Section \ref{sec:hawc}). To do so, we calculate: 
\begin{equation}
    A_{\text{eff}}=\epsilon \pi A_{\text{thrown}}\,,
\end{equation}
where $A_{\text{thrown}}$ is the area over which simulated EASs are thrown and $\epsilon$ is the fraction of events thrown that triggered the detector.
In Figure \ref{fig:areaon}(a), we compare the FHit Pass 4 and Pass 5 analyses as a function of the true energy of the simulated primary gamma ray for events that passed trigger and quality conditions only. In Figure \ref{fig:areaon}(b), we show the subset of events from Figure \ref{fig:areaon}(a) that passed gamma/hadron separation cuts, which naturally decrease the effective area.

Pass 5 has approximately three to five times more effective area at low energies for the range from 0 to 46$^\circ$ zenith mainly due to the inclusion of small events as a result of the
improvement in noise suppression with the MPF algorithm described above (see Section \ref{subsec:lowe}). These events have a relatively poor angular resolution and a high background, so their inclusion does not improve the sensitivity of HAWC to sources extending well above 1 TeV, such as the Crab. 

In Figure \ref{fig:areaon}(a), it is clear that in Pass 4 the area decreased above 100 TeV, regardless of the shower inclination. Pass 5 improves the high energy effective area by over a factor of two, reaching the physical area of the primary array. As described in Section \ref{subsec:apulse}, this was achieved by allowing more PMTs to be affected in large showers by the after-pulsing veto, as it was shown that only having 80\% of PMTs available for reconstruction was not an impediment to improving the performance of the reconstruction algorithms. 

As described in Sections \ref{subsec:dir} and \ref{subsec:ghcuts}, and shown in Section \ref{subsubsection:gheff}, the inclusion of a LDF fit to the EAS footprint for gamma/hadron separation cuts in Pass 5 and its optimization by zenith angle (Section \ref{subsec:cutopt}) improved the background rejection by up to a factor of four for large zenith angles. The impact of this is seen in Figure \ref{fig:areaon}(b) for zenith angles 37$^\circ$ to 46$^\circ$, in which the effective area can be seen to have increased at all energies, reaching the physical area of the primary array at approximately 200 TeV, increasing the chances of detecting a source at a high zenith angle up to energies of $\sim300$~TeV. 

\begin{figure}
\centering
\resizebox{1.0\textwidth}{!}{%
\begin{tabular}{cc}
\includegraphics[width=0.5\textwidth]{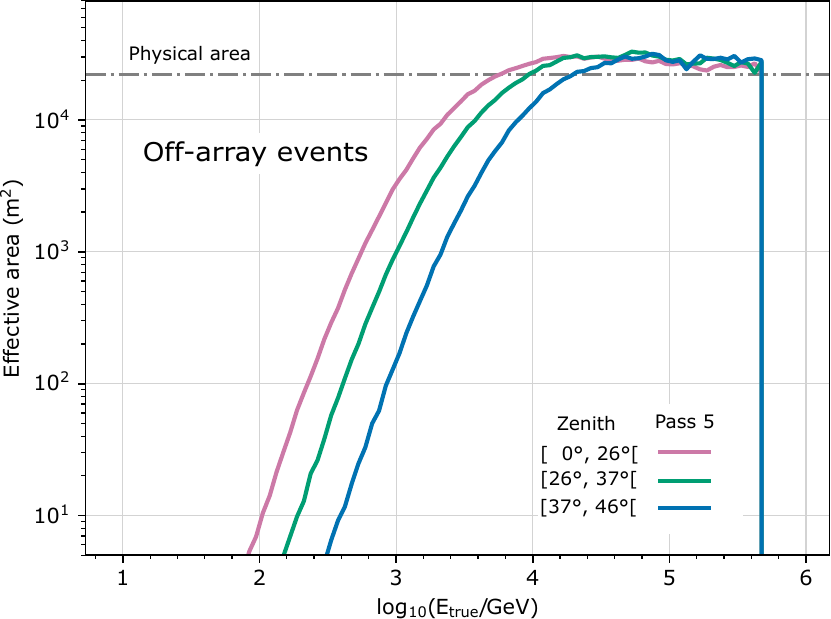}%
&\quad
\includegraphics[width=0.5\textwidth]{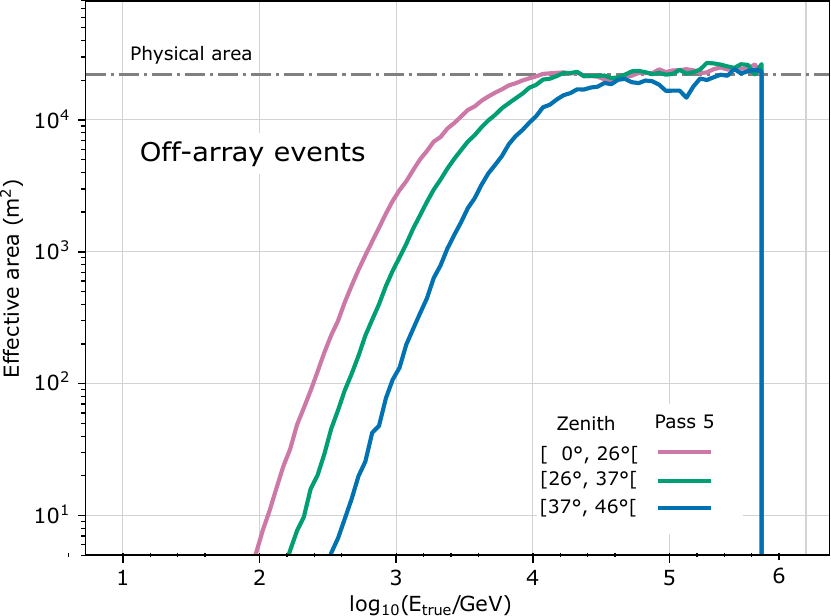}%
\\
(a) Trigger conditions & (b) Gamma/hadron cuts\\
\end{tabular}
}
\caption{Effective area of the Pass 5 EAS reconstruction for three different zenith-angle intervals for \textbf{(a)} events in which the shower core lands off the array and which pass the trigger conditions and \textbf{(b)} events that additionally pass FHit gamma/hadron cuts. We stop plotting the effective area at 500 TeV---as for on-array events---because at higher energy the PMTs receive charge beyond their calibration limit (hardware saturation). No comparison to Pass 4 is shown as we did not bin off-array events separately (see Section \ref{subsec:core}).\label{fig:areaoff}}
\end{figure}

In Figure \ref{fig:areaoff}(a), we show the effective collection area of events with cores landing off-array (see Section \ref{subsec:core}). Figure \ref{fig:areaoff}(b) shows events in (a) that also passed gamma/hadron separation cuts.

In Figure \ref{fig:areaoff}(a), the effective area can be seen to reach $\sim$33,000 m$^{2}$ (physical area $\times 1.5$) at 10 TeV, but the effective area threshold at low energy is larger compared to on-array Pass 5 events for all three zenith angle bins. Off-array events are reconstructed based on the hits of PMTs in the primary array, therefore, low-energy (small) showers may not trigger enough PMTs to be reconstructed. On the other hand, when the core of a large shower lands off-array, the primary array receives mostly the tail of the shower, which does not trigger the after-pulse veto. Therefore, the LDF gamma/hadron separation fit performs better in this part of the shower and the effective area at the highest energies is higher for off-array events than for on-array ones in Pass 5.
After the gamma/hadron separation cuts, the effective area reaches the physical one of the primary array for all three zenith-angle intervals at the highest energies (see Figure \ref{fig:areaoff}(b)).

%%%%%%%%%%%%%%%%%%%%%%%%%%%%%%%%%%%%%%%%%%%%%%%%%%%%%%%%%%%%%5

\subsection{Angular resolution}\label{subsubsection:angle}
\begin{figure}
\centering
\resizebox{1.0\textwidth}{!}{%
\begin{tabular}{cc}
\includegraphics[width=0.5\textwidth]{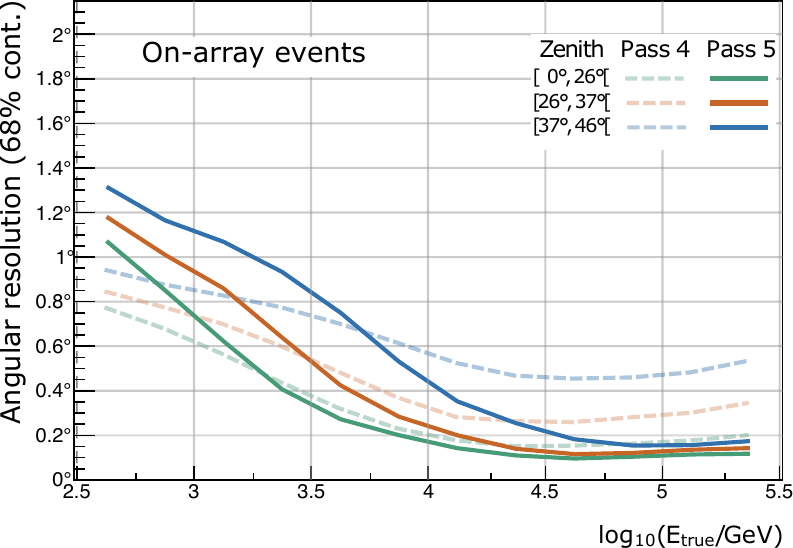}%
&\quad
\includegraphics[width=0.5\textwidth]{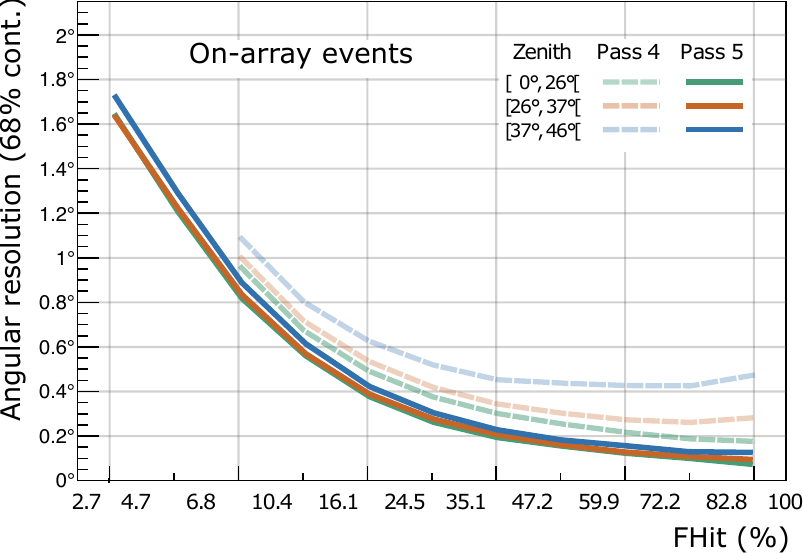}%
\\
(a) Energy & (b) FHit\\
\end{tabular}
}
\caption{Comparison of the angular resolution for on-array events by the Pass 4 and Pass 5 EAS reconstructions for \textbf{(a)} angular resolution as a function of true gamma-ray primary energy---using the NN analysis approach---and \textbf{(b)} angular resolution as a function of FHit. The x-axis shows the lower boundaries of FHit bins as in Table \ref{tab:bins}. See Section \ref{subsubsection:angle} for details. The Pass 4 performance is not the full Pass 4 performance, only of on-array Pass 4 events.\label{fig:angon}}
\end{figure}

%"no dependence on the zenith angle bin" -> maybe we shouldn't say NO dependence, just little dependence or less or something.
%%%%%%%%%%%%%%%%%%%%%%%%%%%%%%%%%%%%%%%%%%%%%%%%%%%%%%%%%%%%%5
Here we define the angular resolution as the difference between reconstructed and true angle containing 68\% of the events. In Figure \ref{fig:angon}(a), we present a comparison of angular resolution as a function of the true gamma-ray primary energy for three zenith-angle intervals for events with shower cores landing on the primary array, using the NN gamma/hadron separation cuts. For each energy bin, the angular distribution was calculated and the value containing 68\% of the events was assigned to the average energy of the bin. Figure \ref{fig:angon}(b) shows the angular resolution as a function of FHit in the same three zenith-angle intervals. We used the same method to calculate the lines. Note that the Pass 5 curves start at lower FHit values due to the inclusion of smaller events, which also causes the worsening of angular resolution at lower energies in panel (a).

The difference between the curves in Figure \ref{fig:angon}(a) illustrates that for gamma rays with the same energy, the zenith angle will determine the energy loss of the shower while traversing the atmosphere, therefore determining the number of secondary particles reaching the detector. The greater the zenith angle of the shower, the poorer the statistics to reconstruct its direction. It makes sense that the best performance is for nearly overhead sources. 

In Pass 4, we reported the 68\% of angular containment for an overhead source as 0.39$^\circ$ for bin B4 and 0.17$^\circ$ for bin B9, with median energies of 2 and 35 TeV, respectively. In Figure \ref{fig:angon}(b), we show that the angular containment is now 0.33 and 0.15$^\circ$ (the corresponding B5 and B10 in Pass 5, on-array events), respectively, not only for overhead sources but for ones located out to 46$^{\circ}$ zenith, where the angular resolution is improved by almost a factor of three. An angular resolution of less than 0.2$^{\circ}$ above 40~TeV is reached for all three zenith-angle ranges. Therefore, not only have we improved the angular resolution of highly inclined sources, but the high zenith angular resolution is also very close to the best performance achievable. The same effect was shown for gamma/hadron separation efficiency in Section \ref{subsubsection:gheff}.

\begin{figure}
\centering
\resizebox{1.0\textwidth}{!}{%
\begin{tabular}{cc}
\includegraphics[width=0.5\textwidth]{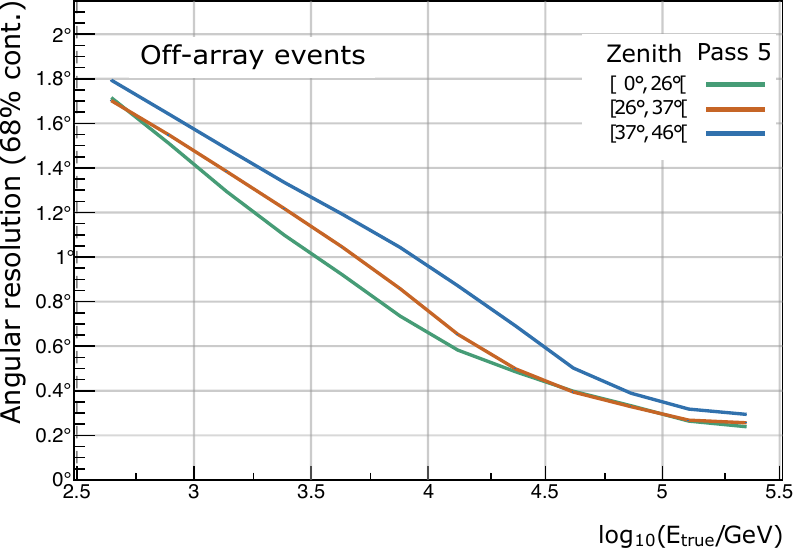}%
&\quad
\includegraphics[width=0.5\textwidth]{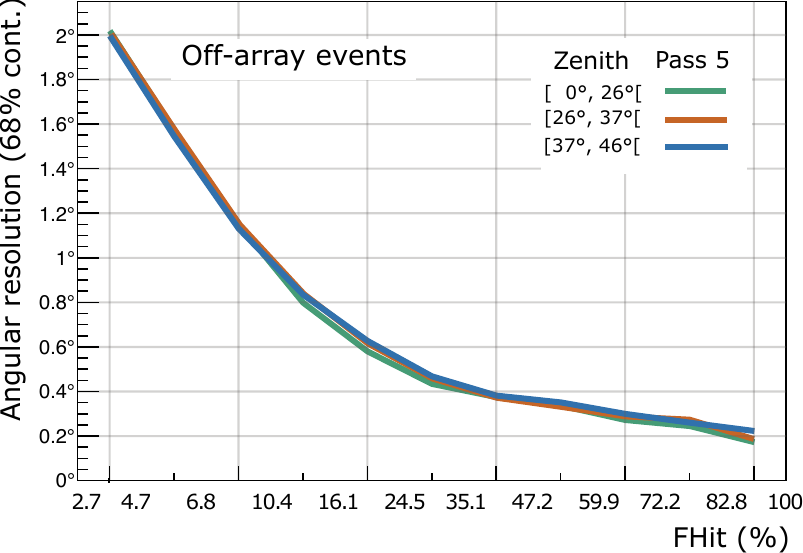}%
\\
(a) Energy & (b) FHit\\
\end{tabular}
}
\caption{Angular resolution achieved by Pass 5 EAS reconstruction of off-array events for \textbf{(a)} angular resolution as a function of true gamma-ray primary energy---using the NN analysis approach---and \textbf{(b)} angular resolution as a function of FHit. The x-axis shows the boundaries of FHit bins as in Table \ref{tab:bins}. Note that no comparison can be made as Pass 4 could not identify off-array events (see Section \ref{subsec:core}).\label{fig:angoff}}
\end{figure}

In Figure \ref{fig:angoff}(a) and (b) we show the angular resolution of events reconstructed with cores landing off the array as a function of true energy---using the NN analysis approach---and as a function of FHit, respectively. When compared to Figure \ref{fig:angon}(a), as a function of energy, it is evident that the angular resolution is worse than for on-array events. In particular, high-zenith-angle events have much worse angular resolution as there are significantly fewer hits landing on the array to reconstruct the shower. Moreover, gamma/hadron separation cuts for off-array events are not optimized in energy bins (see Sections \ref{subsec:ghcuts} and \ref{subsec:cutopt}), therefore, it is expected that the angular resolution does not improve as fast with increasing energy. As a function of FHit, we can also see that off-array events have worse angular resolution than on-array (Figure \ref{fig:angon}(b)), but we still get an agreement between the three zenith angle bins.

\subsection{Gamma/hadron separation efficiency}\label{subsubsection:gheff}

Here we compare the performance of the FHit
gamma/hadron separation cuts in Pass 4 and Pass 5 for on-array events. As explained in Section \ref{subsec:ghcuts}, in Pass 5, we replaced the parameter PINC with the reduced $\chi^{2}$ obtained from a fit to the LDF measured in the HAWC primary array based on the NKG function. We calculate the separation efficiency as the ratio of the
number of events passing the gamma-hadron cuts to the total number of events that passed trigger conditions only. To compute the efficiency of EASs from gamma-ray primaries retained after cuts, $\epsilon_{\gamma}$, we use simulations. For hadronic primaries, $\epsilon_{h}$, we use data, since the vast majority of EASs triggering ground-based
observatories are background events initiated by hadronic primaries (mostly protons).

\begin{figure}[ht!]
% \begin{center}
\centering
\includegraphics[width=0.333\textwidth]{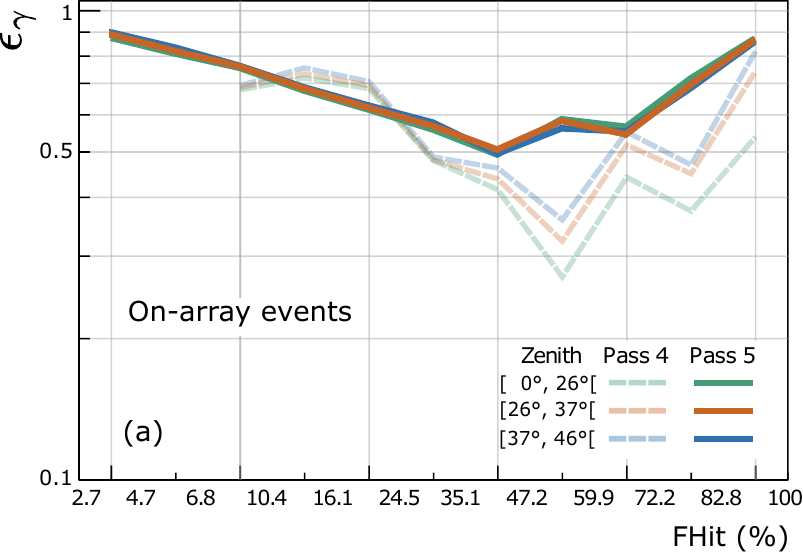}%
\includegraphics[width=0.333\textwidth]{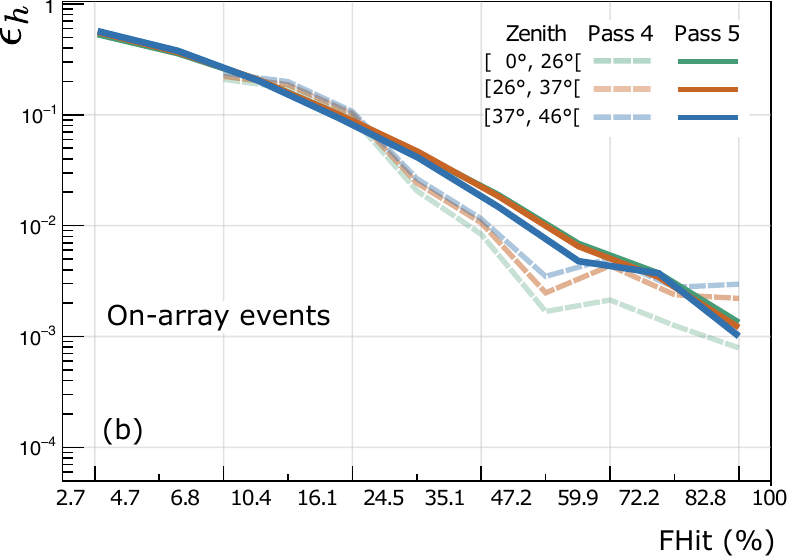}%
\includegraphics[width=0.333\textwidth]{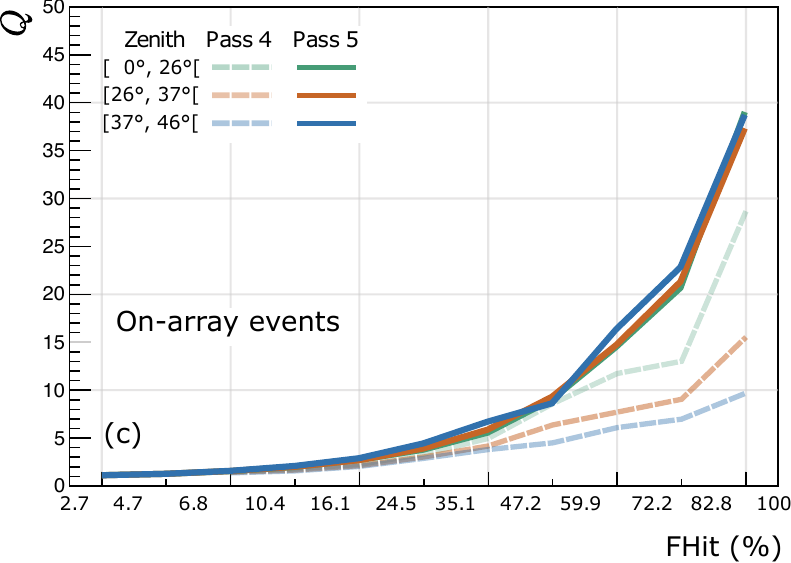}% 
\caption{Comparison (on-array) between the Pass 4 and Pass 5 versions of FHit analysis: \textbf{(a)} gamma/hadron cut efficiency for gamma-ray showers, \textbf{(b)} gamma/hadron cut efficiency for cosmic-ray showers, and \textbf{(c)} Q factor, all plotted as a function of FHit. The x-axis shows the boundaries of FHit bins as in Table \ref{tab:bins}. The Pass 4 performance is not the full Pass 4 performance, only of on-array Pass 4 events.\label{fig:ghon}}
% \end{center}
\end{figure}

In Figure \ref{fig:ghon}(a), we compare the calculated gamma-ray efficiencies $\epsilon_{\gamma}$ in the Pass 4 and Pass 5 FHit reconstructions as a function of FHit. We can see from this that Pass 5 improved the retention of gamma-ray events to $\geq$50\% for all bins. Figure \ref{fig:ghon}(b) shows the same comparison for the hadron efficiencies, $\epsilon_{h}$. As expected, the fraction of events passing the cuts diminishes with the increase in shower size since large showers have more hits than smaller (lower-energy) showers.

Using the hadron and gamma efficiencies we can quantify the increase in Gaussian significance (due to the selection of events by applying the cuts) using the quality factor, or $Q$ factor, which may be defined as:
 
\begin{equation}\label{eq:q}
    Q=\frac{\epsilon_{\gamma}}{\sqrt{\epsilon_{h}}}\,.
\end{equation}

In Figure \ref{fig:ghon}(c), we show the $Q$ factor values vs. FHit for three zenith angle intervals, comparing Pass 4 and Pass 5 HAWC reconstructions. For Pass 5, the $Q$ factor value stays above 1 for all FHit bins. For 24.5\% of PMTs hit and above, it is evident that gamma/hadron separation cuts were improved for all zenith-angle intervals. The most significant enhancement is at the highest zenith-angle interval, where the quality factor improvement is up to a factor of 4 for the most inclined showers. Furthermore, Figures \ref{fig:ghon}(a), (b), and (c) show that there is almost no dependence on the zenith angle for $\epsilon_{\gamma}$, $\epsilon_{h}$ and $Q$ with respect to FHit, as was the case for the angular resolution performance seen in Figure 4(b).

\begin{figure}[ht!]
% \begin{center}
\centering
\includegraphics[width=0.333\textwidth]{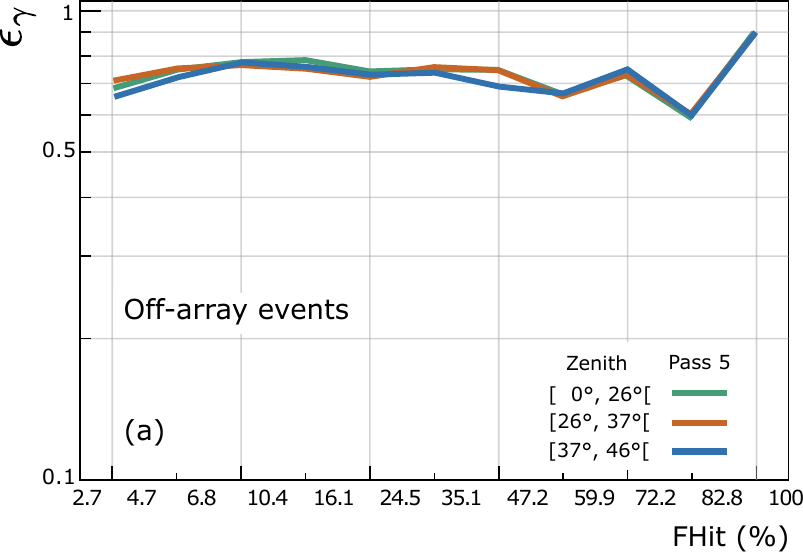}%
\includegraphics[width=0.333\textwidth]{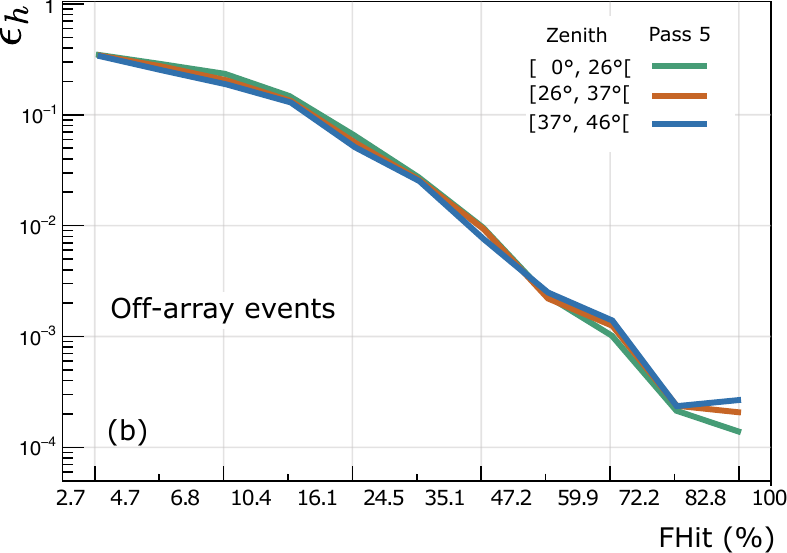}%
\includegraphics[width=0.333\textwidth]{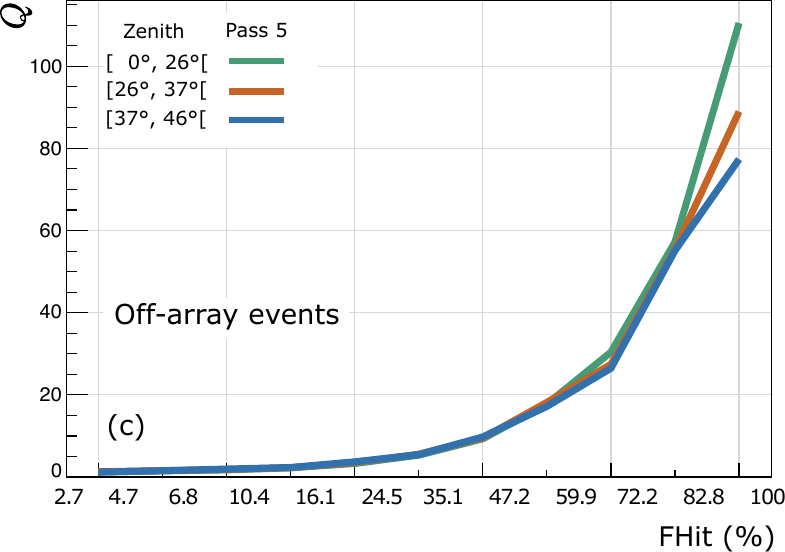}% 
\caption{Off-array events in the Pass 5 version: \textbf{(a)} Gamma/hadron cut efficiency for gamma-ray showers, \textbf{(b)} gamma/hadron cut efficiency for cosmic-ray showers, and \textbf{(c)} Q factor, all plotted as a function of FHit. The x-axis shows the boundaries of FHit bins as in Table \ref{tab:bins}.  \label{fig:ghoff}}
% \end{center}
\end{figure}

In addition, we show in Figure \ref{fig:ghoff}(a) $\epsilon_{\gamma}$, (b) $\epsilon_{h}$, and (c) $Q$ for off-array events as a function of FHit. The overall hadron rejection performance is better than on-array events, unlike for angular resolution. Gamma-ray retention is over 60\% for all bins and the $Q$ factor increases for all three zenith-angle intervals in the last FHit bin (82.2\%--100\%) where, unlike on-array events, the zenith-angle intervals reach different values. For the first two bins (2.7\%--6.8\%) in Pass 5, $\epsilon_{\gamma}$ is between 80\% and 90\% for on-array events, whereas for off-array events $\epsilon_{\gamma}$ is 70\%-80\%. This difference between showers with cores landing on- and off-array is expected. For a small fraction of PMTs hit, the efficiency of the gamma/hadron separation depends on the fit to the lateral distribution of hits and the maximum charge measured by a PMT outside a radius of 40 m around the core location (see Section \ref{subsec:ghcuts}). Thus, it makes sense that small gamma-ray induced showers with cores landing on the primary array are easier to distinguish. On the other hand, the efficiency for large high-energy showers is worse for on-array events because the particles in the core are energetic enough to trigger the after-pulsing veto described in Section \ref{subsec:apulse}. In contrast, with the same large fraction of PMTs hit, events with cores landing off array are less likely to trigger the veto and still provide enough information to fit the lateral distribution.

\section{Verification with observation of the Crab Nebula}\label{sec:crab}

We test and validate the reconstruction algorithms by observing the Crab Nebula. This gamma-ray source is stable, well measured, and possesses a bright and ``steady'' emission, making cross-instrument performance comparisons and calibrations easier.

\subsection{Data set and angular resolution}\label{subsec:data}

The following results include events coming from the vicinity of the Crab, which culminates just 3$^\circ$ from zenith for HAWC. The data are reconstructed as described in Section \ref{sec:rec}, and events that pass the optimized selection discussed in Section \ref{subsec:cutopt} are divided into a set of maps, one for each FHit or reconstructed energy bin. The latter contain the number of signal and background events---calculated with the direct integration method \citep{atkins2003observation,abdo2012observation}---according to the FHit values of the reconstructed EAS events. 

\begin{figure}
\begin{center}
\resizebox{1.0\textwidth}{!}{%
\begin{tabular}{ccc}
\includegraphics[width=0.33\textwidth]{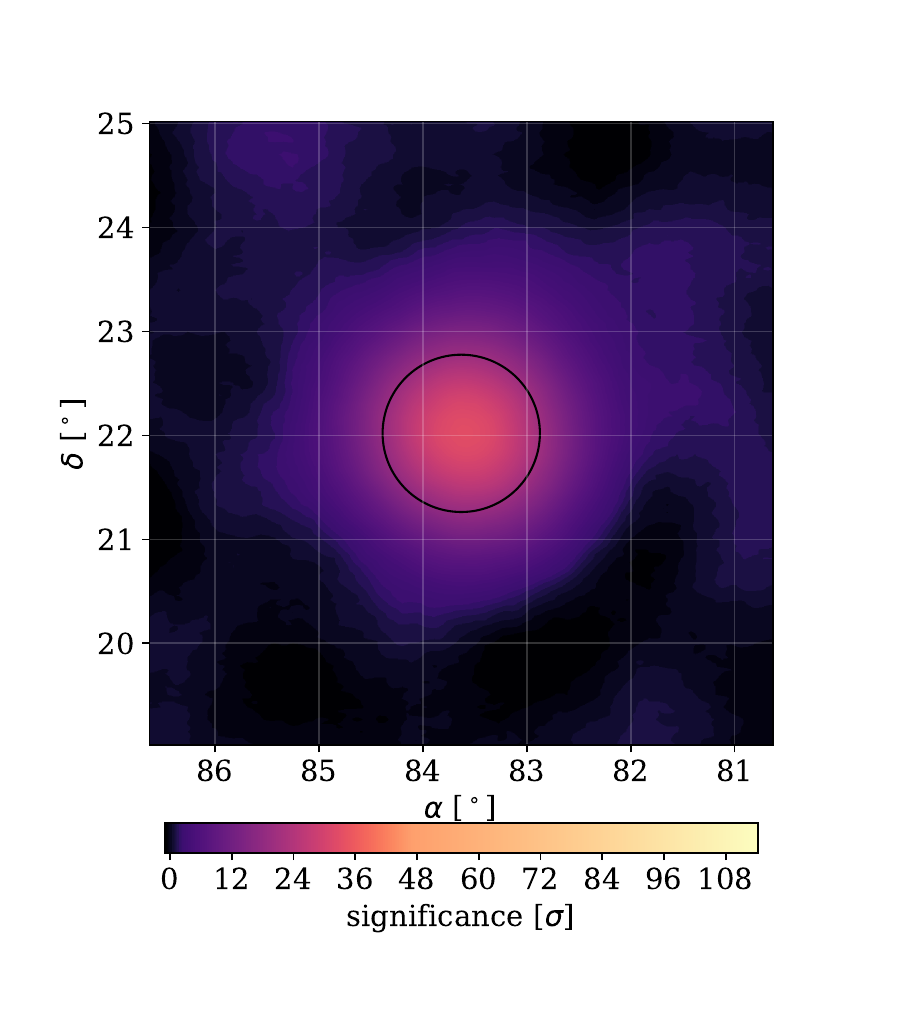}%
&\quad
\includegraphics[width=0.33\textwidth]{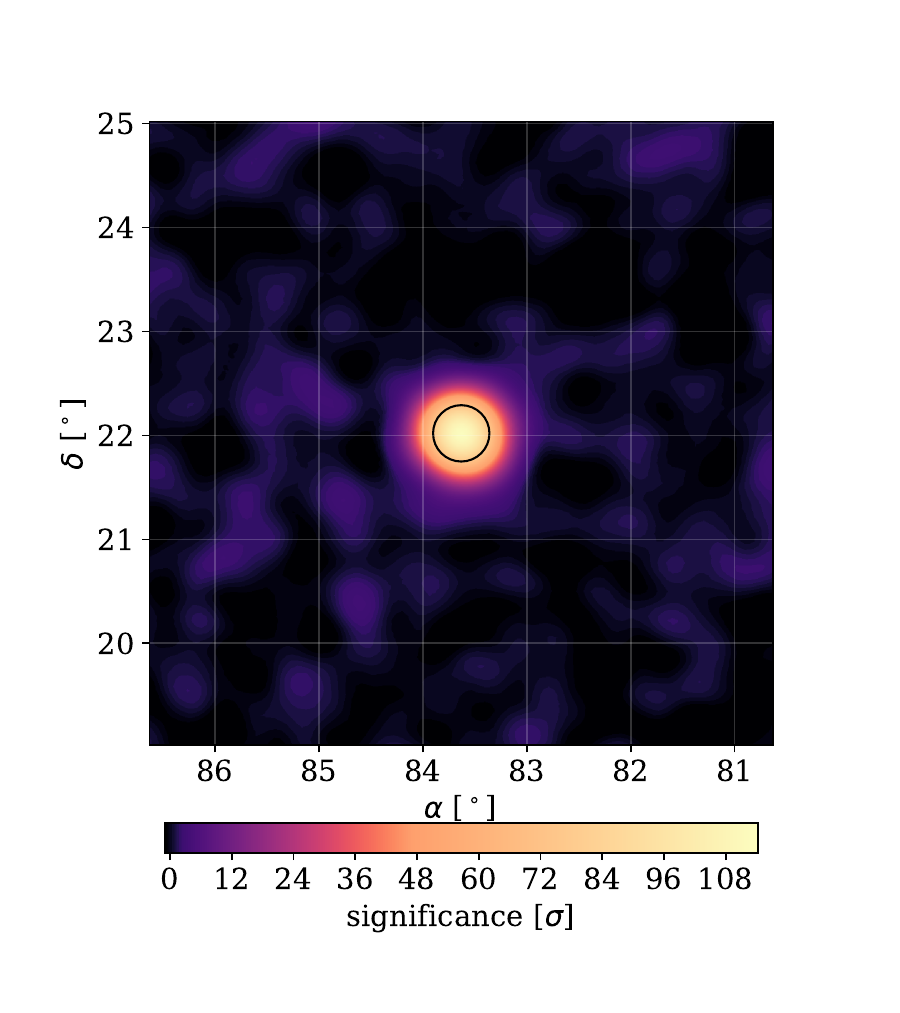}%
&\quad
\includegraphics[width=0.33\textwidth]{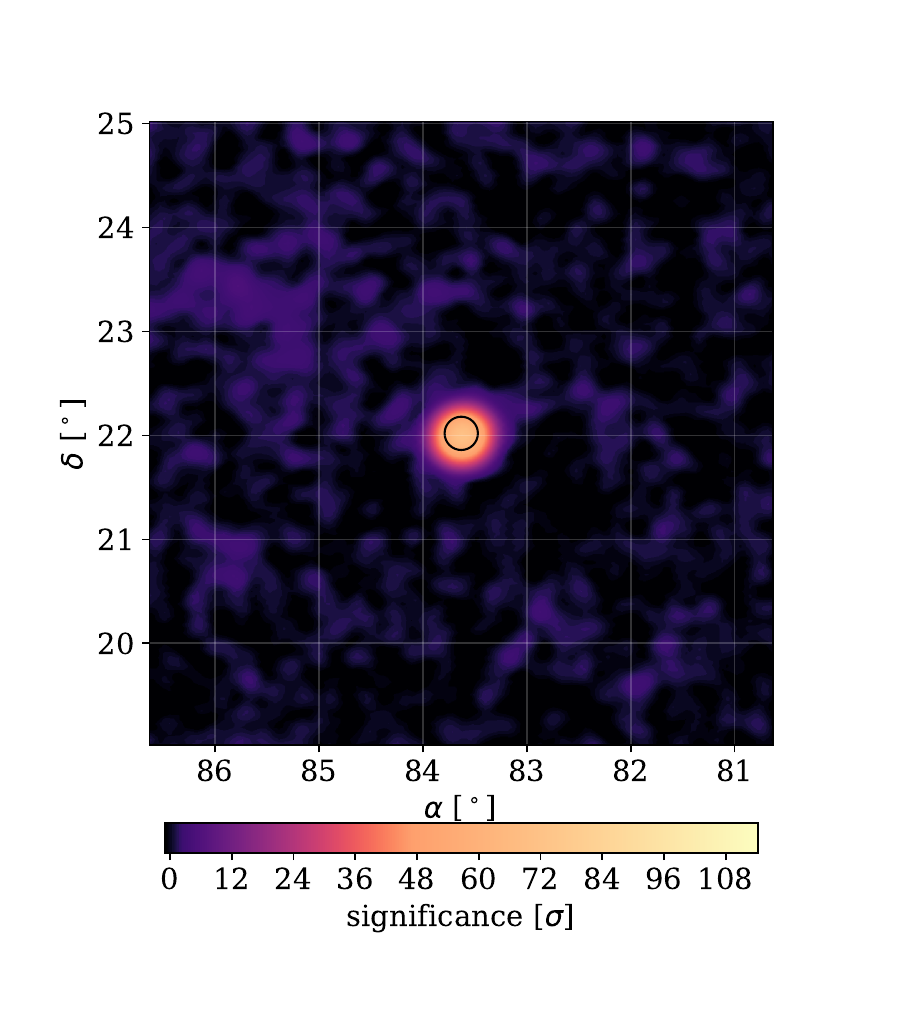}%
\\
B2 on-array & B6 on-array &B10 on-array\\
\includegraphics[width=0.33\textwidth]{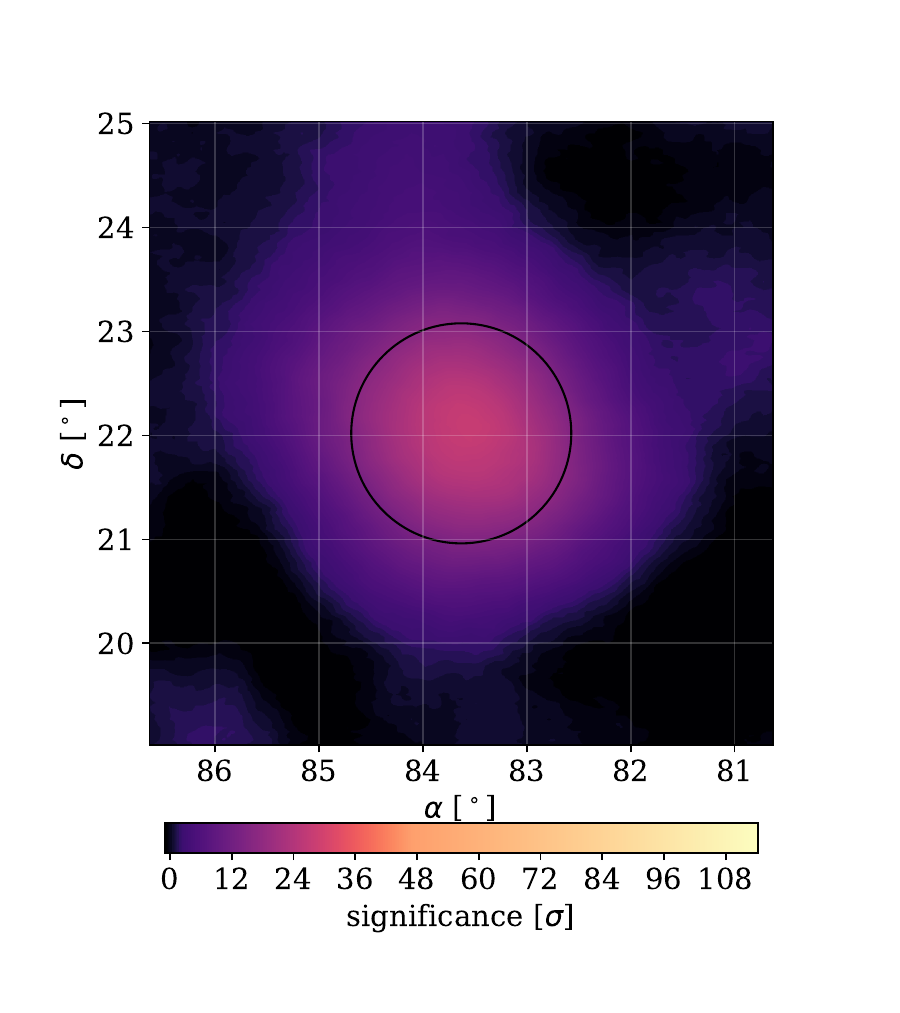}%
&\quad
\includegraphics[width=0.33\textwidth]{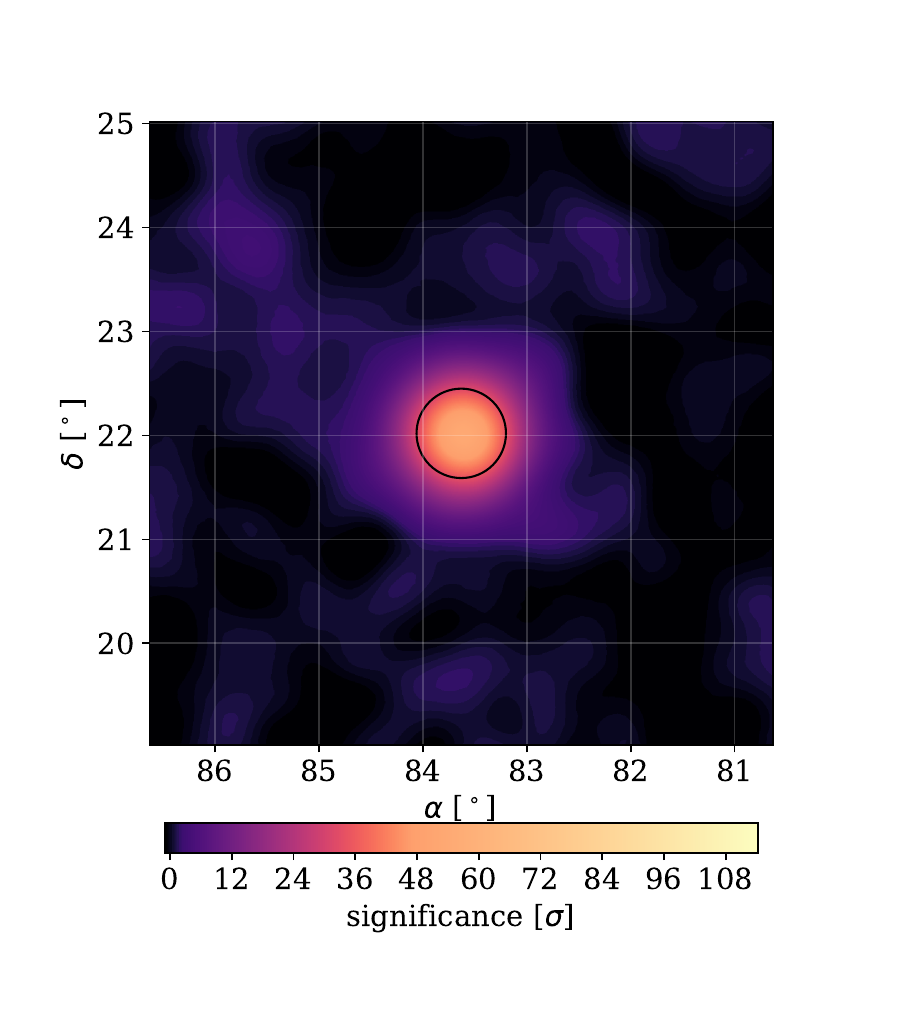}%
&\quad
\includegraphics[width=0.33\textwidth]{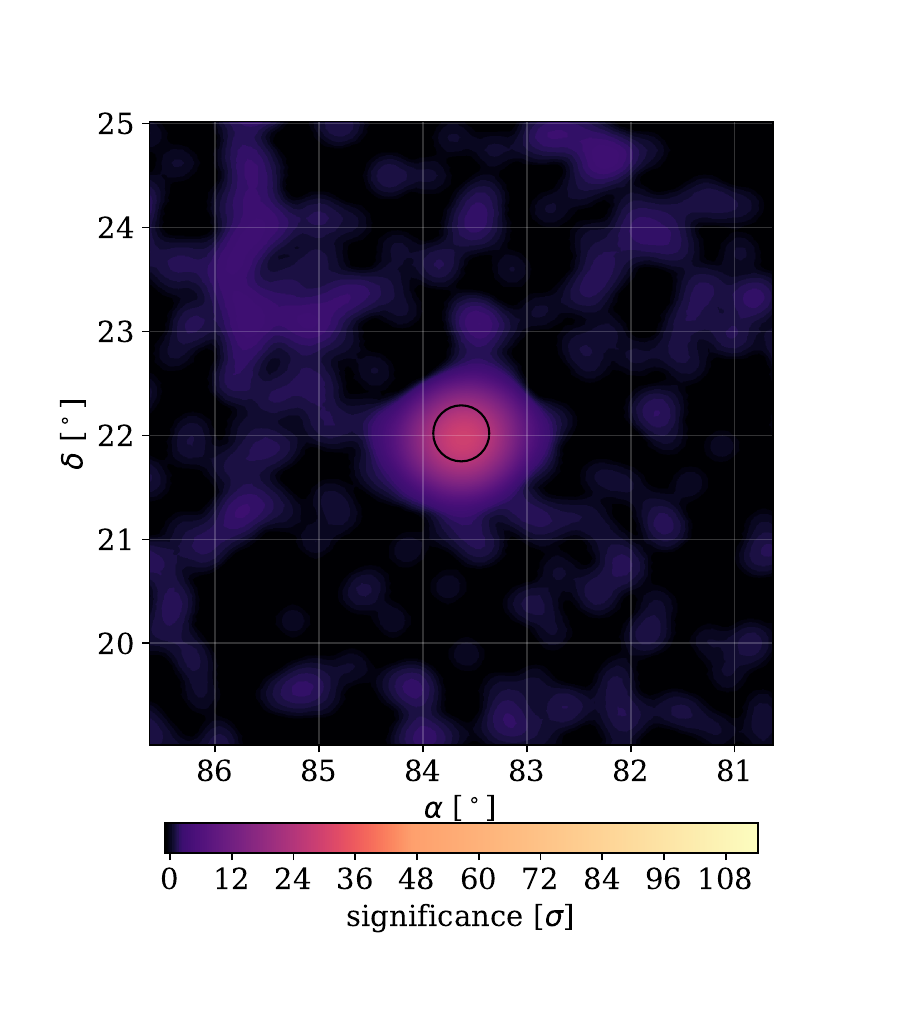}%
\\
B2 off-array & B6 off-array  &B10 off-array \\
\end{tabular}
}
\caption{Significance maps of the selected bins. The columns correspond to FHit bin B2 (6.8--10.4\%), bin B6 (35.1--47.2\%), and bin B10 (82.8--100\%). The first row corresponds to on-array events, and the second to off-array events. The black circles show the area of 68\% of containment from Figure \ref{fig:resfrommaps}. We calculate the significance approximated as the square root of the TS, under the validity of the Wilks' theorem \citep{wilks1938large}. We calculate the significance approximated as the square root of the log likelihood ratio of signal to background, using the Likelihood Fitting Framework (LIFF) \citep{younk2015high,abeysekara2017observation}. See Table \ref{tab:bins} for median energies, Table \ref{tab:crabsig} for maximum significance, and Section \ref{subsec:data} for details about the data. \label{fig:sigmaps}}
\end{center}
\end{figure}

During the 2565 days of data used for the analysis presented here, we observed the Crab Nebula with a significance of 258$\sigma$ using on-array events, and 123$\sigma$ with off-array events. As shown in Figures \ref{fig:angon} and \ref{fig:angoff}, the angular resolution of the HAWC reconstruction is strongly correlated to FHit. Thus, we show the performance using FHit bins (see Table \ref{tab:bins}) rather than energy intervals. To make the significance maps in Figure \ref{fig:sigmaps}, we selected events from a region of $\pm 3^{\circ}$ from the Crab Nebula in FHit bin B2 (6.8--10.4\% of the PMTs hit), bin B6 (35.1--47.2\%), and bin B10 (82.8--100\%), for on and off-array events. These bins represent small, medium and large showers, respectively. The maximum significance for these bins are summarized in Table \ref{tab:crabsig} and the median energies in Table \ref{tab:bins}. The maximum significance detection comes from bin B6 (see Figures \ref{fig:sigmaps}(b) and (e), on and off array respectively). From these significance maps one can already tell that the off-array reconstruction results in a wider radial distribution of events (worse angular resolution) and lower significance.

\begin{table}
    \centering
    \begin{tabular}{|c|c|c|}
    \hline
       Bin  &  Maximum significance ($\sigma$)\\
         \hline
       B2 on-array &  33.0\\
           B2 off-array  & 27.8\\
         \hline
       B6 on-array  & 114.2\\
            B6 off-array  & 54.1\\
         \hline
       B10 on-array & 71.6\\
            B10 off-array  & 29.2\\
         \hline
    \end{tabular}
    \caption{Summary of maximum significance of events from the Crab Nebula, for the selected bins.}
    \label{tab:crabsig}
\end{table}

\begin{figure}
\begin{center}
\resizebox{1.0\textwidth}{!}{%
\begin{tabular}{ccc}
\includegraphics[width=0.33\textwidth]{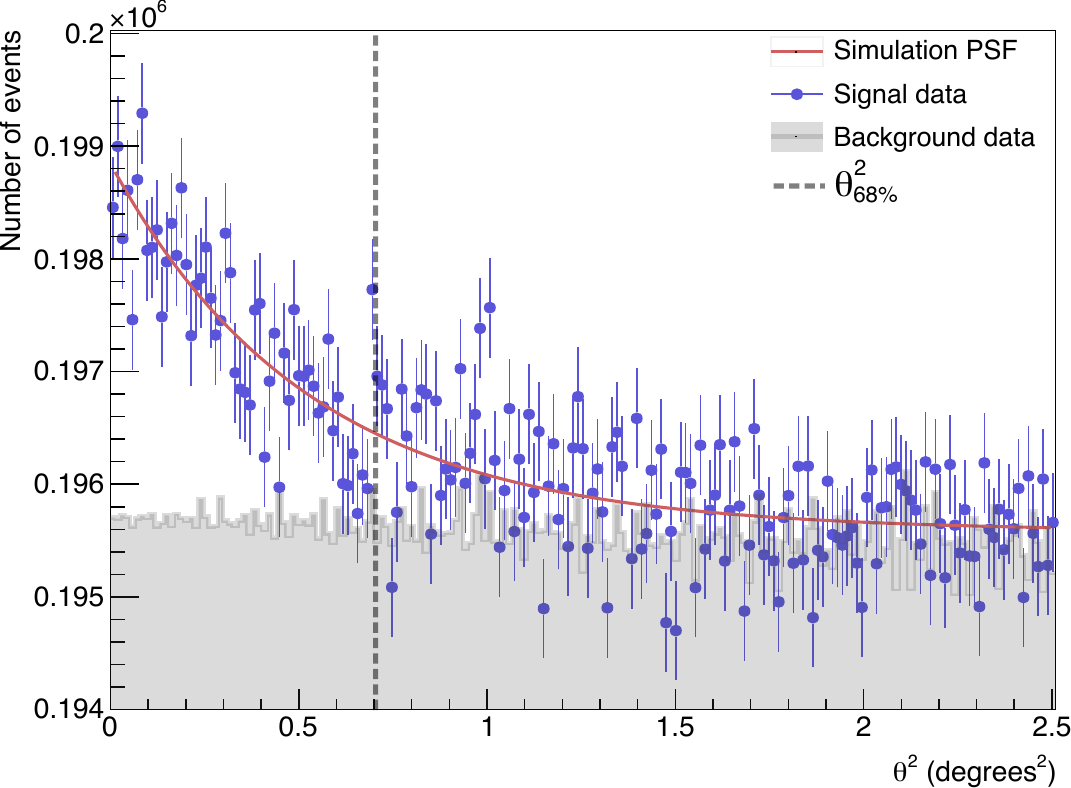}%
&\quad
\includegraphics[width=0.33\textwidth]{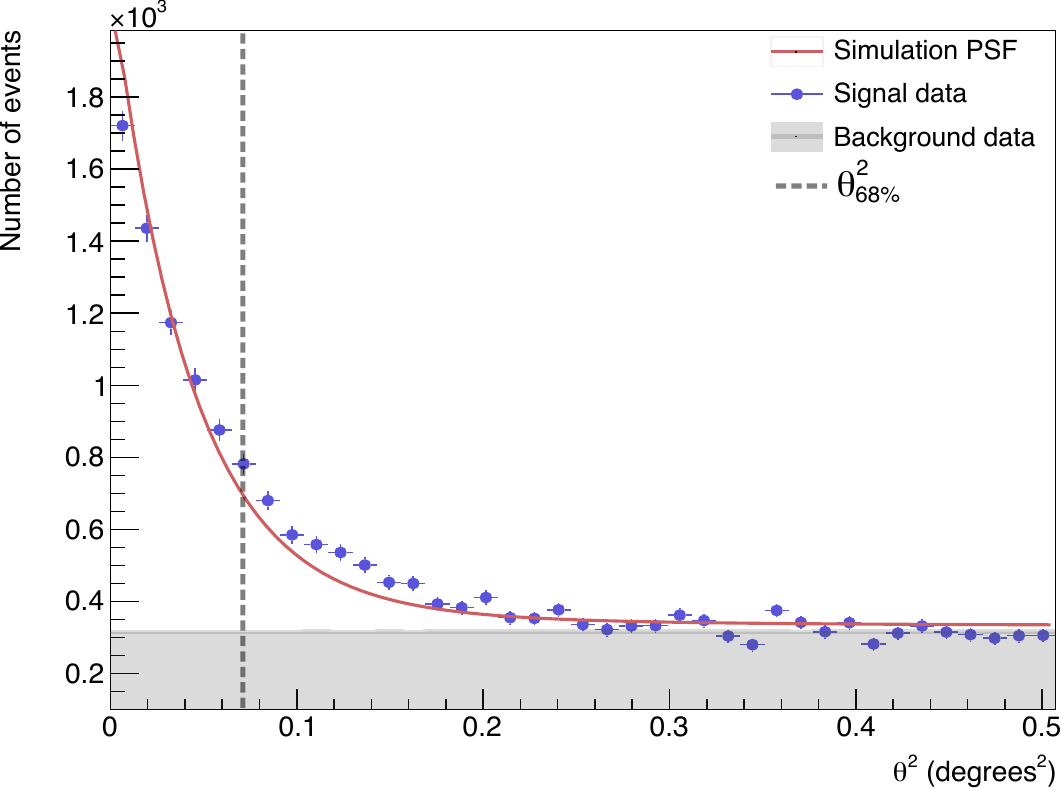}%
&\quad
\includegraphics[width=0.33\textwidth]{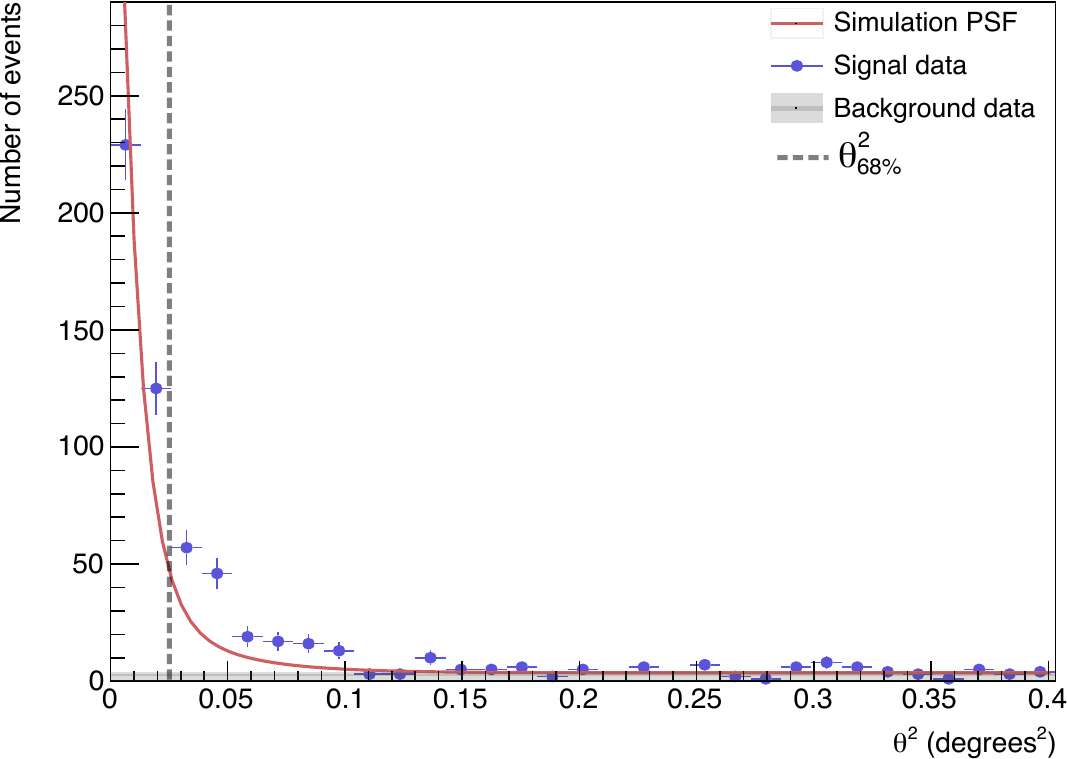}%
\\
B2 on-array ($\theta_{\text{68\%}}=0.836^{\circ}$) & B6 on-array ($\theta_{\text{68\%}}=0.271^{\circ}$) &B10 on-array ($\theta_{\text{68\%}}=0.160^{\circ}$)\\
\includegraphics[width=0.33\textwidth]{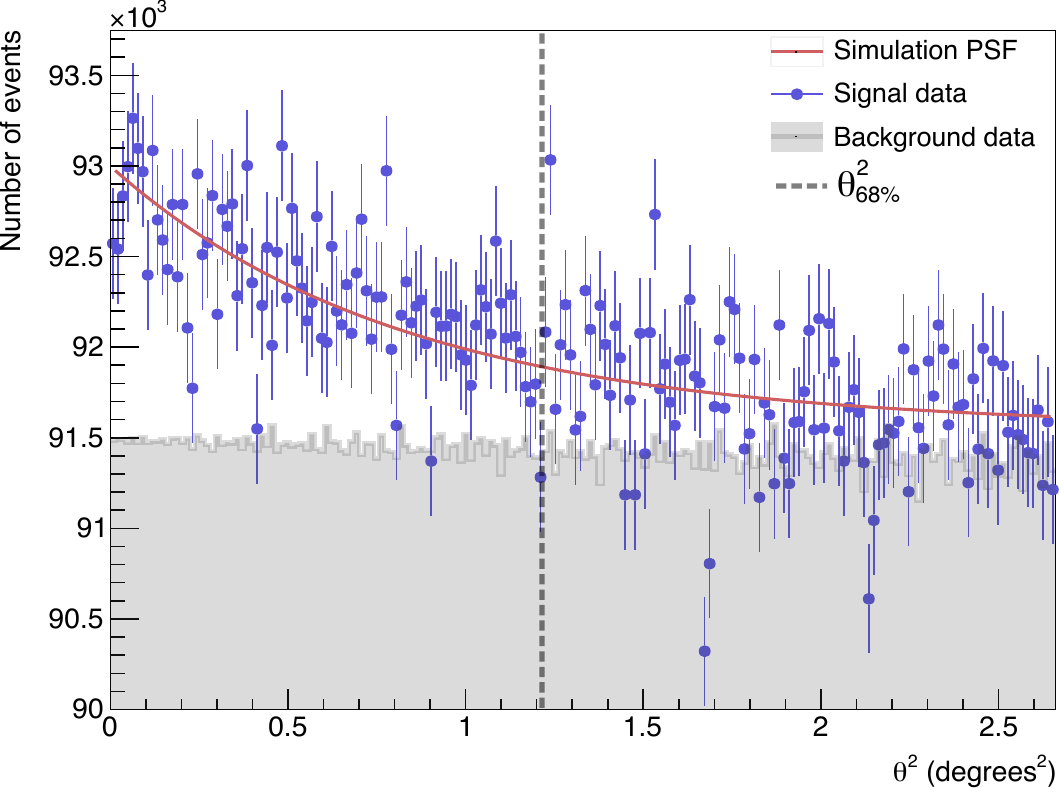}%
&\quad
\includegraphics[width=0.33\textwidth]{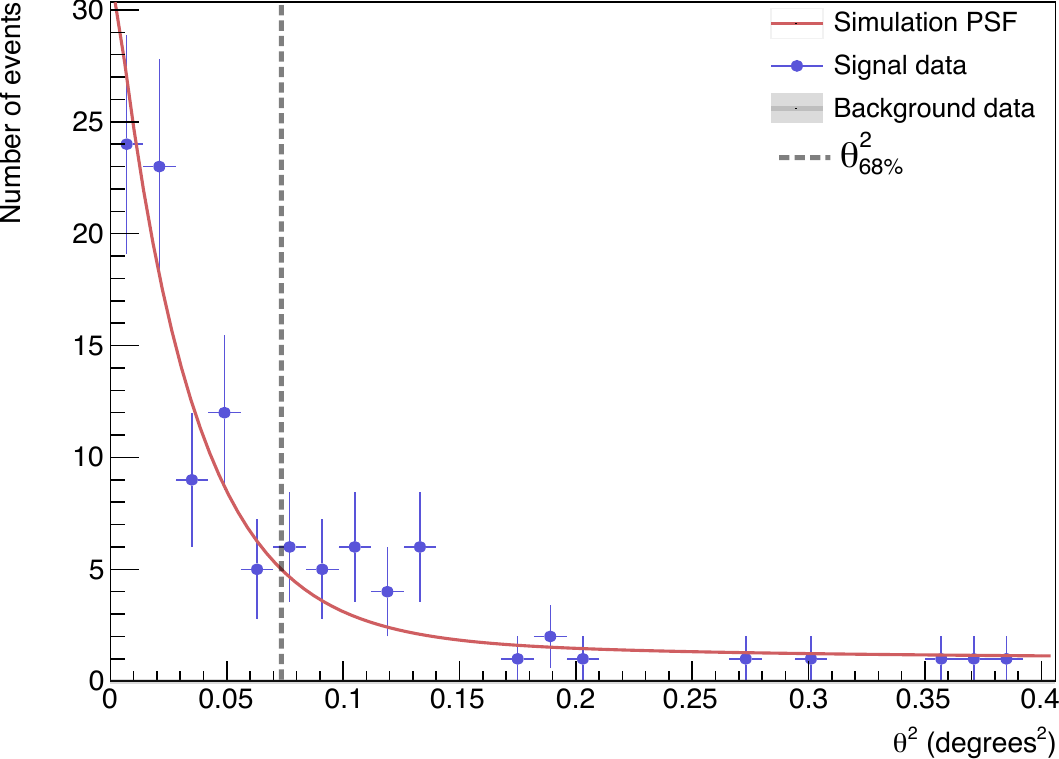}%
&\quad
\includegraphics[width=0.33\textwidth]{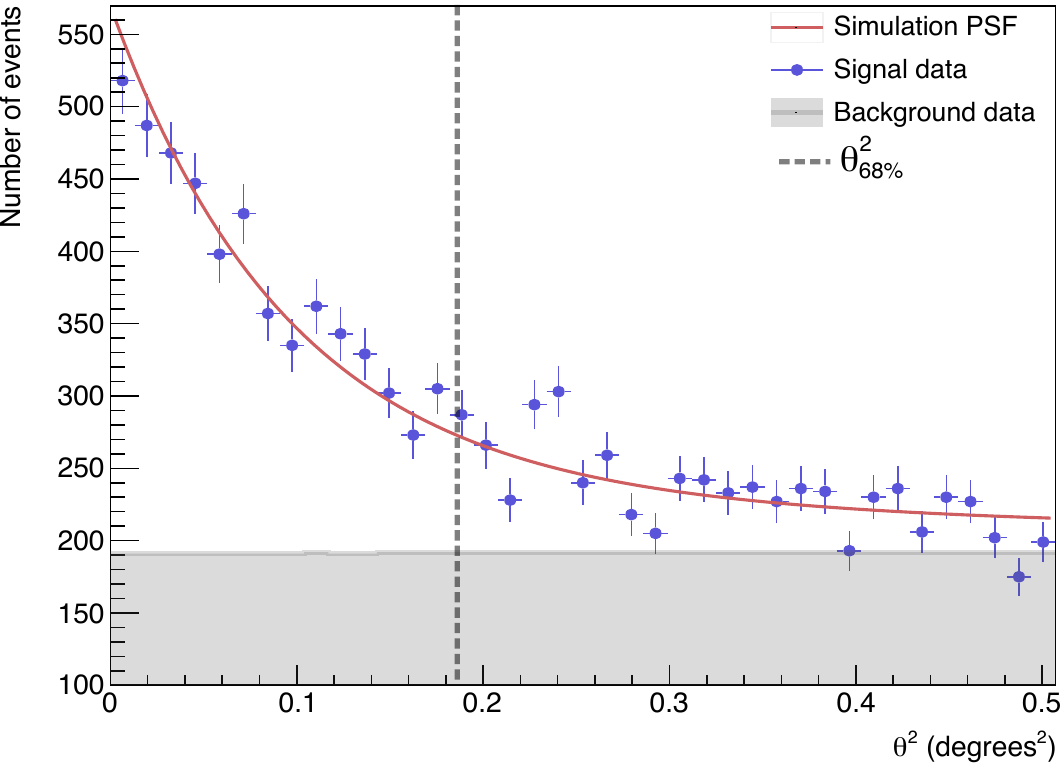}%
\\
B2 off-array ($\theta_{\text{68\%}}=1.059^{\circ}$) & B6 off-array ($\theta_{\text{68\%}}=0.430^{\circ}$) &B10 off-array ($\theta_{\text{68\%}}=0.269^{\circ}$)\\
\end{tabular}
}
\caption{On- and off-array event distributions for chosen bins. These distributions are ordered as follows: the columns correspond to FHit bin B2 (6.8--10.4\%), bin B6 (35.1--47.2\%), and bin B10 (82.8--100\%). The first row corresponds to on-array events, and the second, to off-array events. For each sub-figure we indicate the angle of 68\% of containment ($\theta_{\text{68\%}}$). Note that the axes are different with a larger number of events available for this study at lower FHit values. See Tables \ref{tab:bins} and \ref{tab:crabsig} for median energies and maximum significance, respectively, and Section \ref{subsec:data} for details about the data.\label{fig:resfrommaps}}
\end{center}
\end{figure}

In Figure \ref{fig:resfrommaps}, we show the distribution of signal and background events---as selected with the gamma/hadron separation cuts described in Section \ref{subsec:ghcuts}---from the Crab Nebula location for all three selected bins as a function of $\theta^{2}$, where $\theta$ is the measured angle relative to the Crab Nebula's true location. The point spread function (PSF) of HAWC is well described by:
\begin{equation}\label{eq:PSF}
    \text{PSF}(\psi) = CG_{1}(\psi)+(1-C)G_{2}(\psi)\,,
\end{equation}
where $G_{i}$ are Gaussians of different widths, $\psi$ is the angular error in the reconstruction, and $C$ is a fit parameter as described in \citep{abeysekara2017observation}.
Here we use data to verify that the PSF described in equation \ref{eq:PSF} is well simulated for small, medium, and large events for on and off-array events. As expected from the simulation performance plots in Figures \ref{fig:angon} and \ref{fig:angoff}, the signal distribution becomes narrower as the shower size increases, which improves the statistics for better performance of the reconstruction algorithms. The differences in $\theta_{\text{68\%}}$, the angle of 68\% of containment, between simulations and data range from 0.01$^{\circ}$ to 0.06$^{\circ}$. As expected, on-array events have narrower distributions than off-array events, and the uncertainties are larger for the latter (see Figures \ref{fig:angon} and \ref{fig:angoff}). For small events the efficiency of the gamma/hadron separation is low but the number of events is high, causing large fluctuations in the data. Since the size of the map pixels ($\sim 0.008^{\circ}$ side) is significantly smaller than the angular resolution of the most precise bin in the data, we expect some misplaced events between adjacent pixels.

\subsection{Spectral energy distribution}\label{subsec:sed}

\begin{figure}
\begin{center}
        {\includegraphics[width=0.55\textwidth]{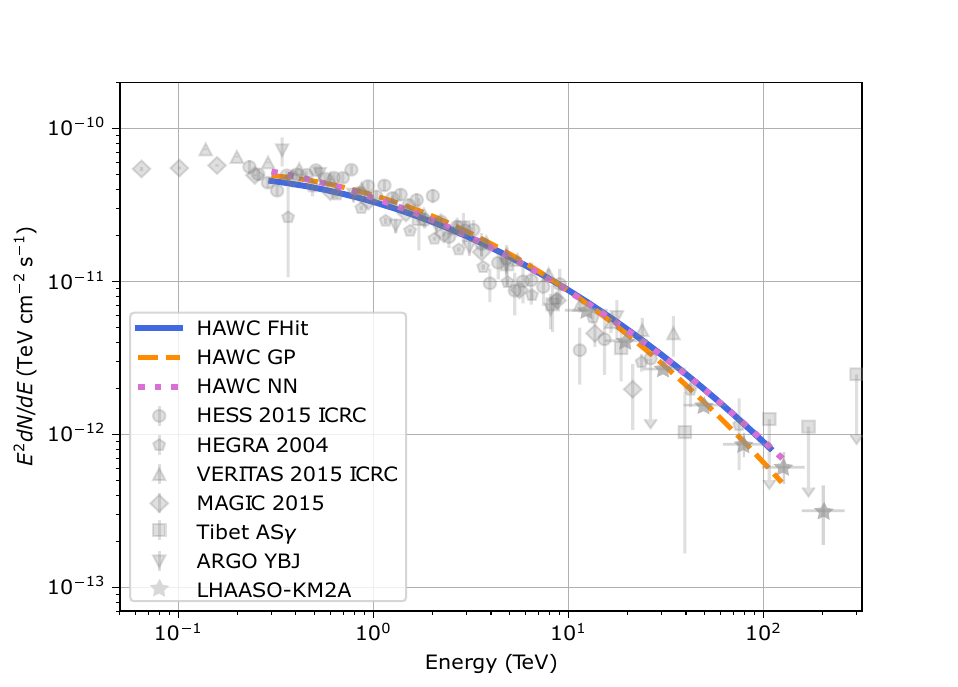}}
    \caption{Spectral energy distribution of HAWC's three analytical approaches with other experiments for comparison. The results for FHit are shown in blue, for GP in orange, and for NN in magenta (see Table \ref{tab:bins}). The HAWC spectra cover an energy range from 300~GeV to the upper energy limits: 108~TeV, 126~TeV, and 125~TeV for FHit, NN, and GP, respectively. Other experiments include H.E.S.S. \citep{holler2015observations}, HEGRA \citep{Aharonian_2004}, VERITAS \citep{meagher2015six}, MAGIC \citep{Aleksi__2015}, Tibet AS-$\gamma$ \citep{amenomori2015search}, ARGO-YBJ \citep{Bartoli_2015}, and LHAASO \citep{aharonian2021observation}. \label{fig:sed}}
    \end{center}
\end{figure}

We performed a maximum likelihood fit to the following log parabola shape
\begin{equation}\label{eq:p}
    \frac{dN}{dE}=\phi_{0}(E/E_{0})^{-\alpha-\beta\ln (E/E_{0})}\,,
\end{equation}
where the free fit parameters are the indices $\alpha$ and $\beta$, and $\phi_{0}$ is the flux normalization \citep{abeysekara2019measurement}. The pivot energy $E_{0}$ is calculated first and then fixed to minimize the correlation between the other parameters for all three analyses (FHit, GP, and NN). We use $E_{0}=2$ TeV in this analysis. In Pass 4, the pivot energy of 7 TeV was used. It decreased, presumably due to the inclusion of lower-energy events. The fit is performed using the Multi-Mission Maximum Likelihood (3ML) framework \citep{Vianello:201667} combined with the HAWC Accelerated Likelihood (HAL) package \citep{abeysekara2022characterizing} applied to the three sets of HAWC sky maps. Table \ref{sedtable} shows the remaining parameter values obtained for all three analyses and their statistical errors. For the fit we are using on-array bins in the three analysis approaches, and also off-array bins in FHit. In Figure \ref{fig:sed} we show a comparison of the best fits obtained for the FHit, GP, and NN approaches with the measurements of other experiments.

\begin{table}[H]
\begin{tabular}{ |c|c|c|c|c| } 
 \hline
 Analysis approach & $\phi_{0}$ $\times 10^{-12} (\text{TeV cm}^{2}\text{ s)}^{-1}$& $\alpha$ & $\beta \times 10^{-1}$ \\ 
 \hline
 Fraction of PMTs hit (FHit)&6.13 $\pm$ 0.04& 2.495 $\pm$ 0.007& 0.89 $\pm$ 0.04 \\ 
   Ground Parameter (GP) & 6.70 $\pm$  0.05& 2.526 $\pm$ 0.007 &1.08 $\pm$  0.04\\ 
    Neural Network (NN)& 6.30 $\pm$ 0.04& 2.537$\pm$  0.007& 0.79 $\pm$  0.04\\ 
 \hline
\end{tabular}
\caption{\label{sedtable}Fit parameter values  and statistical errors obtained for three analysis approaches. The flux normalization $\phi_{0}$ is evaluated at 2 TeV.}
\end{table}

\section{Systematic uncertainties}\label{sec:syst}

Simulation data plays a vital role in the event reconstruction stage. In order to produce realistic simulated data, we need to model the detector as accurately as possible. However, there are some elements of the detector for which there is insufficient understanding and, thus, greater uncertainty in the simulation of them. To account for this fact, we simulate a range of reasonable models and calculate the systematic uncertainties in the same way as in previous HAWC publications \citep{abeysekara2019measurement}. Four systematic uncertainties were found to be non-negligible in the previous energy-dependent study: late-light simulation, uncertainty in the PMT charge-detection threshold, charge uncertainty among PMTs, and absolute PMT efficiency comparing different epochs in time. Previously negligible systematic errors remain
negligible.

\begin{figure}
\centering
\resizebox{0.8\textwidth}{!}{%
\begin{tabular}{cc}
\includegraphics[width=0.4\textwidth]{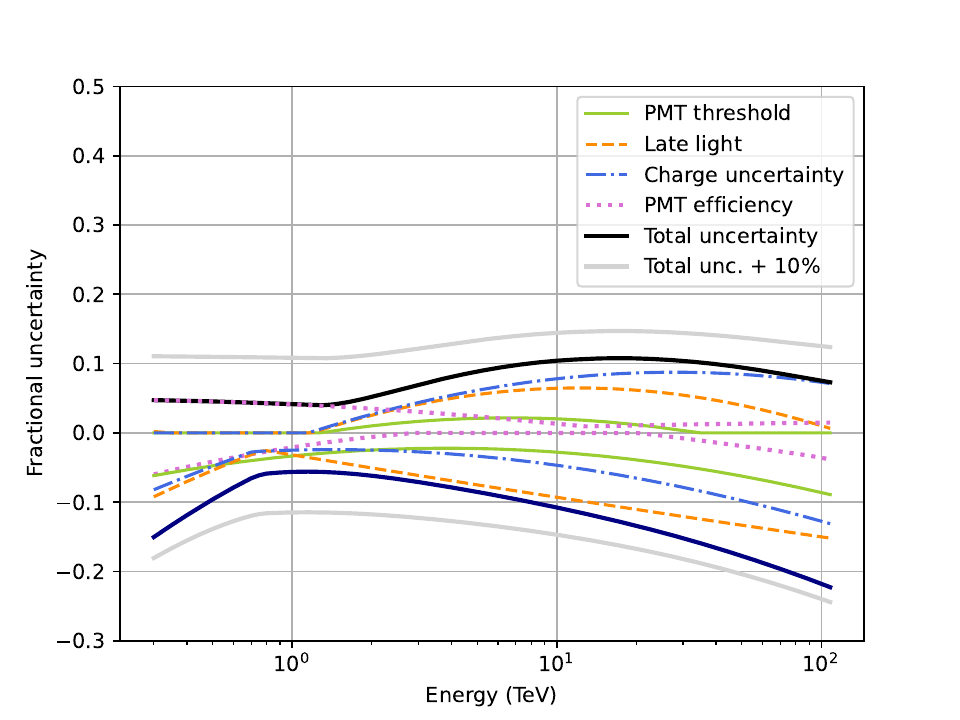}%
&
\includegraphics[width=0.4\textwidth]{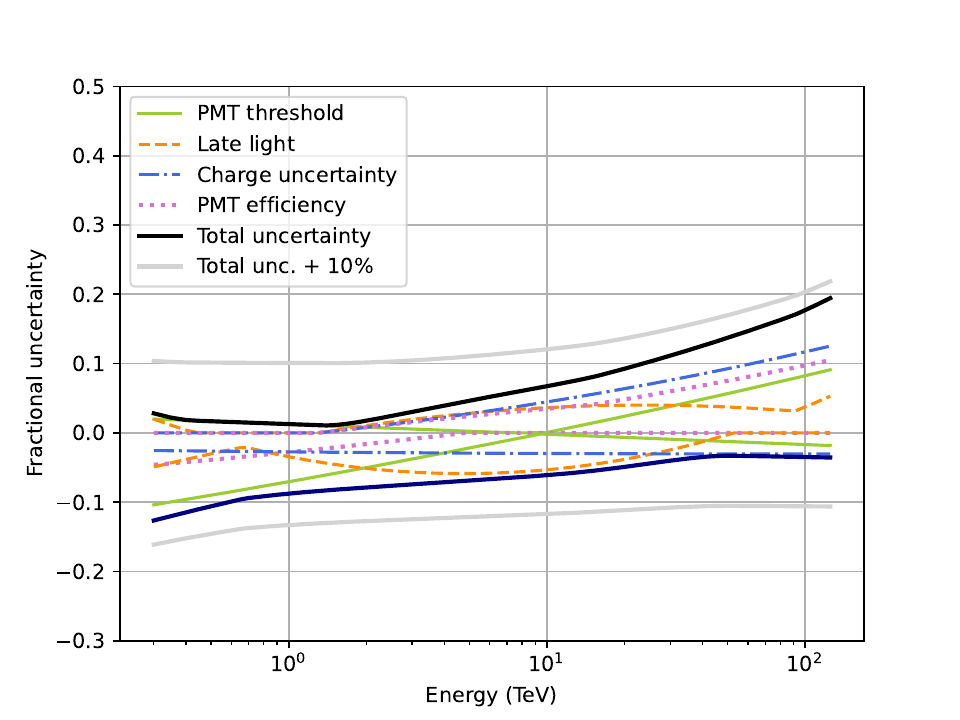}%
\\
(c) FHit (on and off-array)& (b) GP \\
\end{tabular}
}
\\
\centering
\resizebox{0.4\textwidth}{!}{%
\begin{tabular}{c}
\includegraphics[width=0.4\textwidth]{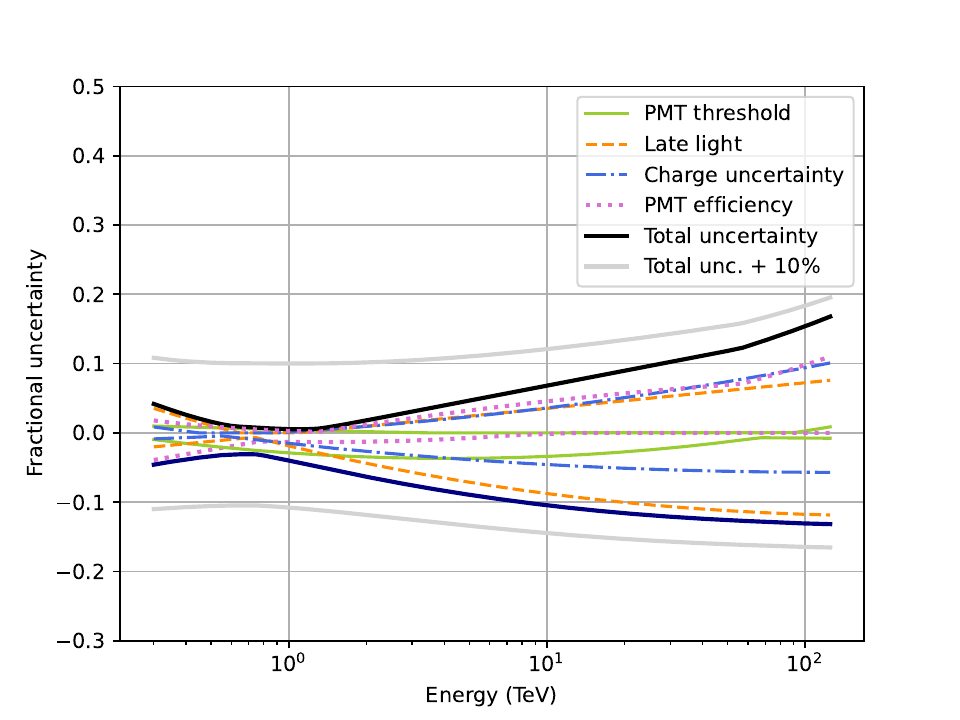}%\\
\\
(c) NN\\
\end{tabular}
}
\caption{Systematic uncertainty extrema of $E^{2}dN/dE$ as a function of energy for the \textbf{(a)} FHit on- and off-array, \textbf{(b)} NN, and \textbf{(c)} GP. We show five different sources of uncertainty: PMT threshold, late light, charge uncertainty, PMT efficiency, and an extra 10\% to account for various minor systematic uncertainty sources. See Section \ref{sec:syst} for details.\label{fig:sys}}
\end{figure}

To estimate the contribution of the systematic uncertainties quoted above and shown in Figure \ref{fig:sys}, we produced instrument response functions (IRFs) with different configurations regarding these uncertainty sources and performed a fit to the Crab Nebula's spectrum. Then, we calculated the fractional uncertainty in $E^{2}dN/dE$ compared to the results shown in Section \ref{subsec:sed}, as a function of energy, always keeping the most significant uncertainty. As in previous papers, the uncertainties are added in quadrature. We also show the total with a conservative extra 10\% of $E^{2}dN/dE$ to account for various minor systematic uncertainties arising from detector and analysis method effects, such as the choice of interaction model in CORSIKA and fluctuations in barometric pressure over time. These variations can cause changes in the detector trigger rate, impacting the rate of hadronic events \citep{abeysekara2019measurement}.

This is the first time we show the systematic uncertainties in an energy-dependent way for FHit. In \citet{abeysekara2017observation}, the overall uncertainty in the flux was quoted as 50\%. However, here we show that the total systematic uncertainty is bounded by $\pm20$\% in all three analysis approaches, for the entirety of the energy range. In comparison to the study shown in \citet{abeysekara2019measurement}, the maximum improvement in the GP and NN analytical approaches are 3\% and 5\% in total fractional uncertainty, respectively, between 1 and 100 TeV.

\section{Sensitivity}\label{section:sensitivity}

\begin{figure}[ht!]
\begin{center}
\resizebox{1.0\textwidth}{!}{%
\begin{tabular}{cccc}
\includegraphics[width=0.5\textwidth]{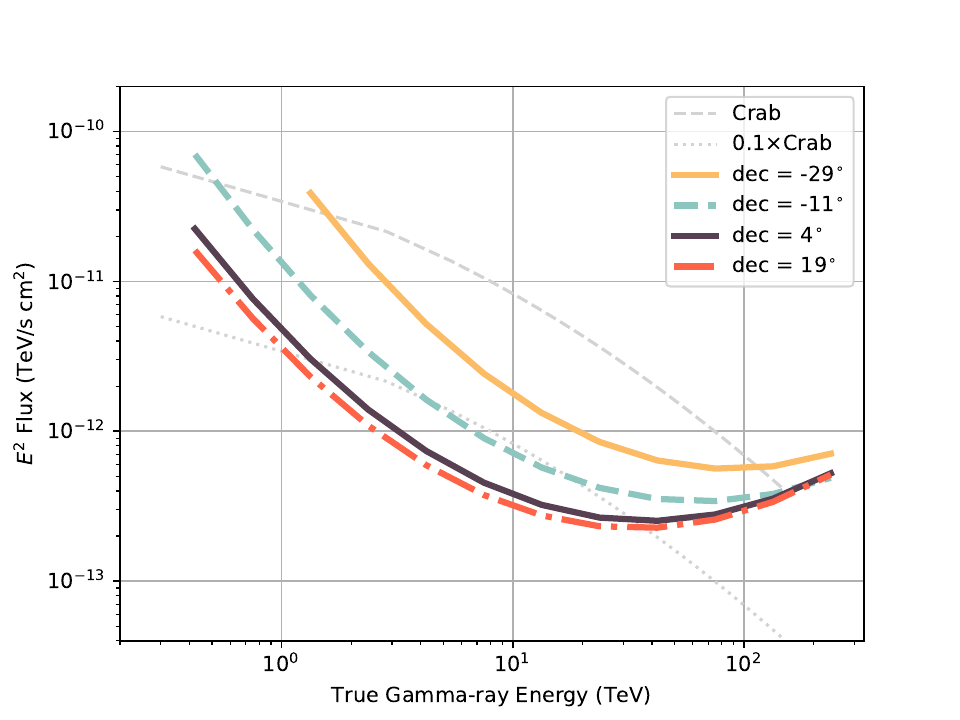}%
&\quad
\includegraphics[width=0.5\textwidth]{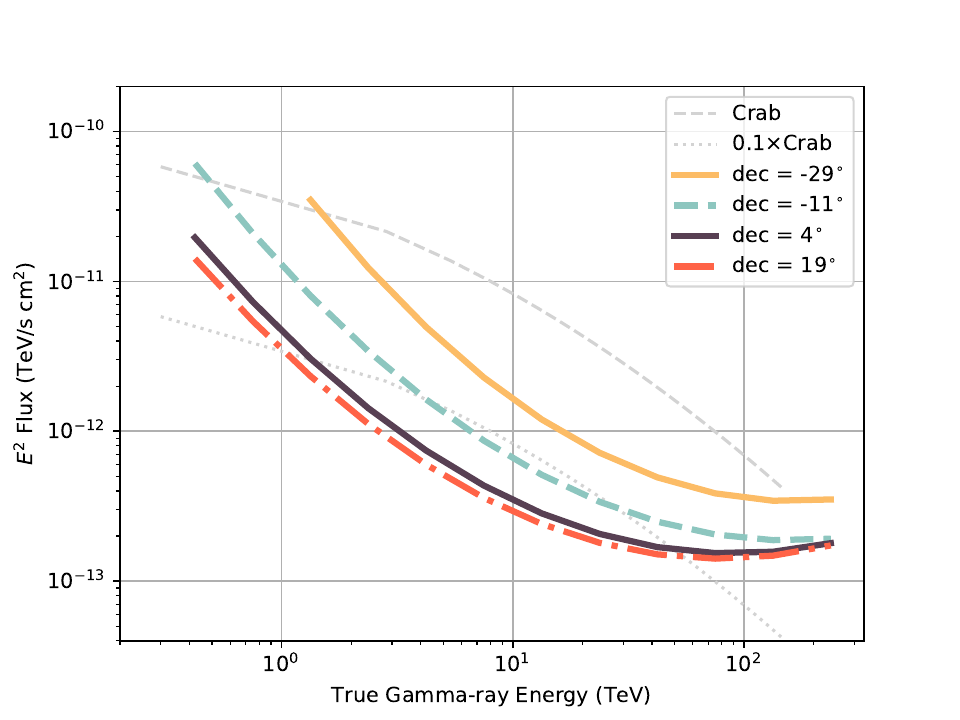}%
\\
(a) FHit on-array & (b) FHit off-array\\
\includegraphics[width=0.5\textwidth]{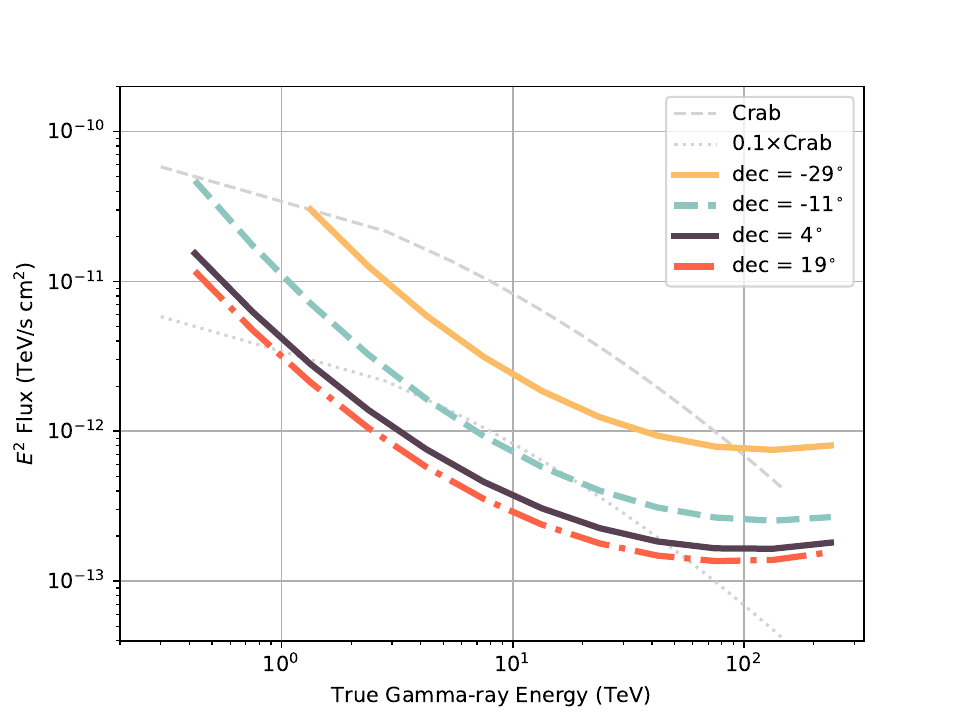}%
&\quad
\includegraphics[width=0.5\textwidth]{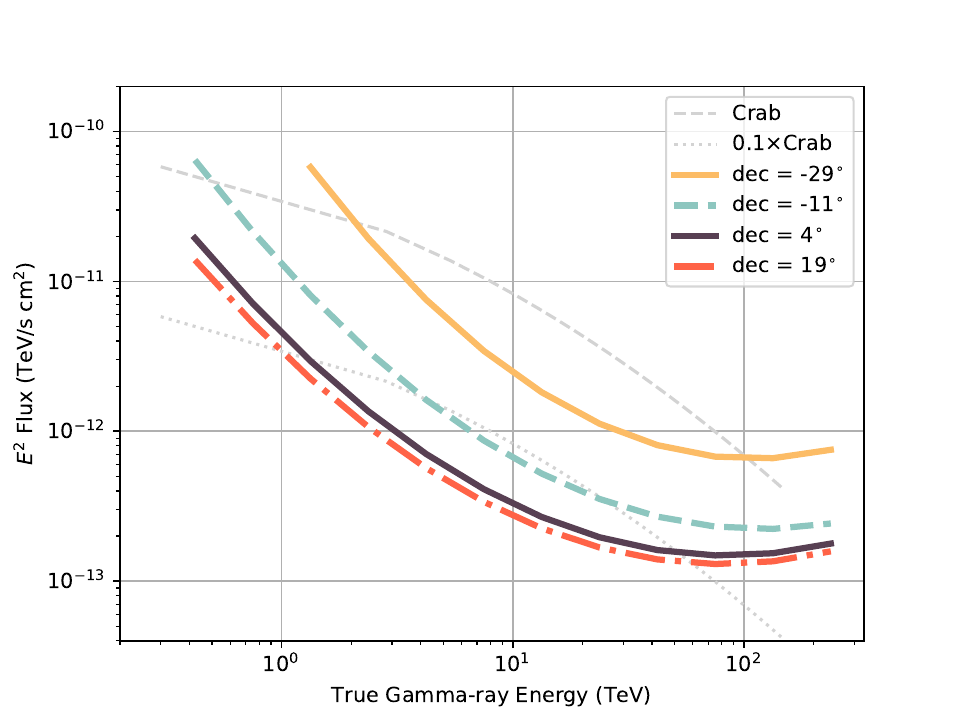}%
\\
(c) GP & (d) NN\\
\end{tabular}
}
\caption{HAWC's quarter-decade-binned differential sensitivity as a function of primary photon's reconstructed energy. \textbf{(a)} FHit on-array events only, \textbf{(b)} FHit on- and off-array events (see Section \ref{subsec:core}), \textbf{(c)} GP, and \textbf{(d)} NN.\label{fig:sens}}
\end{center}
\end{figure}

\begin{figure}
\begin{center}
        {\includegraphics[width=0.6\textwidth]{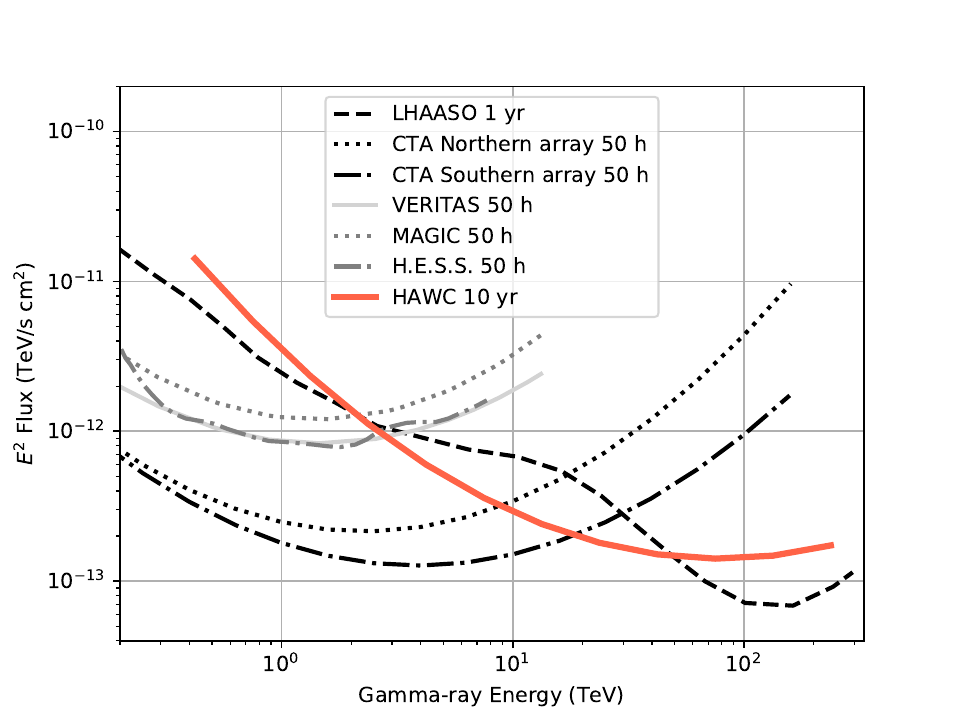}}
    \caption{Comparison of HAWC's sensitivity to other gamma-ray observatories: LHAASO \citep{cao2019large}, CTA Northern and Southern arrays \citep{zanin2022cta}, VERITAS \citep{veritas}, MAGIC \citep{aleksic2016major}, and H.E.S.S. (adaptation by \citet{zanin2022cta}). The comparison of sensitivity between different instruments using the curves shown can only be considered approximate, as the criteria for evaluation and the calculation methods used are not identical. \label{fig:comparison}}
    \end{center}
\end{figure}

In Figure \ref{fig:sens}, we show the HAWC sensitivity to transiting Crab-like point sources (i.e., with a spectral dependence of $E^{-2.63}$ and 2\% of the Crab flux) for FHit (on and off-array events), GP and NN. The lines indicate a flux threshold for a 5$\sigma$ detection, 50\% of the time, with ten years of exposure. The flux values are calculated in true quarter-decade energy bins and then plotted at the median of the energy distribution corresponding to that specific bin and declination. Lastly, we perform a polynomial fit. The resulting sensitivity curves correspond to culminating at zenith angles of 0$^{\circ}$, $15^{\circ}$, $30^{\circ}$, and $48^{\circ}$, which equate to declinations 19$^{\circ}$, 4$^{\circ}$, -11$^{\circ}$ and -29$^{\circ}$. The curves shift to lower sensitivity (i.e. a higher source flux required for detection) and higher energies for declinations farther away from HAWC latitude. In Figures \ref{fig:sens}(a) and (b), we show that the addition of off-array events improves the sensitivity above 10 TeV for all declinations. At 100 TeV, the 5$\sigma$ flux threshold is two times lower when including off-array events. The sensitivity lines for GP and NN are fairly similar (see Figures \ref{fig:sens}(c) and (d)), with flux thresholds comparable to 10\% of the Crab flux between 400 GeV and 100 TeV at 19$^{\circ}$ and 4$^{\circ}$ of declination. Compared to energy estimators, the FHit sensitivity including off-array events is slightly better. We also show a rough comparison of the on- and off-array event sensitivity to other experiments in Figure \ref{fig:comparison}.

\begin{figure}
\begin{center}
        {\includegraphics[width=0.6\textwidth]{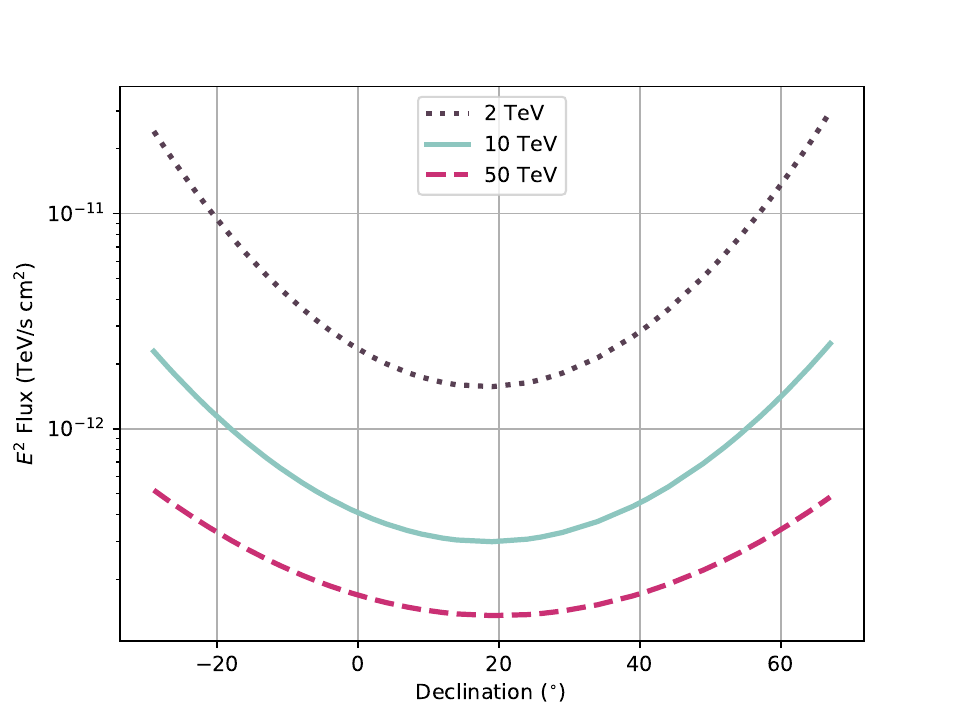}}
    \caption{HAWC's quarter-decade-binned differential sensitivity as a function of declination for three fixed reconstructed energies: 2, 10, and 50 TeV using FHit (on and off-array events, see Section \ref{subsec:core}).\label{fig:sensdec}}
    \end{center}
\end{figure}

In Figure \ref{fig:sensdec}, we show how the sensitivity changes as a function of declination for three energy values: 2, 10, and 50 TeV. These lines are obtained by interpolating the FHit-bin flux thresholds for several declinations and performing a polynomial fit. As discussed above, the sensitivity for all three energies reaches a maximum at HAWC latitude and worsens moving north and south of it. At all declinations, the sensitivity at 2 TeV is much worse than at 10 and 50 TeV.

Finally, due to its wide field of view and long-time exposure, HAWC is also well-suited for observing extended sources. However, the sensitivity---calculated as described above---is expected to be worse than for point sources since more background events are included in a wider region. As a reference, if we compare overhead Crab-like point sources at 1 TeV to extended ones, the flux required for a 5$\sigma$ detection increases by 37\% for a 0.5$^\circ$ wide source and by 103\% for a 1.0$^\circ$ one.

\section{Conclusions} \label{sec:conc}

HAWC has significantly improved its reconstruction algorithms in Pass 5. The effective area is increased by up to 5 times at low energies, which is primarily due to the inclusion of small events thanks to the use of the MPF noise suppression algorithm. These events are significant for our research as they open the opportunity to detect distant active galaxy nuclei and gamma-ray bursts emitting at the energy detection threshold. At the highest energies with Pass 5, the effective area increased and now reaches the primary array's physical area. This was achieved by allowing somewhat more afterpulse self-vetoing of available PMTs for large showers (high energies). Regarding the angular resolution, the most significant improvement is at the highest zenith angles, where the angular resolution has improved by almost a factor of three. The accuracy of HAWC analyses at high gamma-ray energies is much better than previous publications, with an angular resolution of less than 0.2$^{\circ}$ above 40~TeV for overhead to highly inclined sources. This was achieved by correcting biases in direction fitting and biases in highly inclined showers by applying a zenith angle-dependent shift and a minor timing adjustment, along with enhancing core reconstruction by generating PDFs to perform a maximum likelihood fit. The replacement of PINC in the gamma/hadron-separation stage with the reduced $\chi^{2}$---obtained when fitting the LDF with a modified NKG function---reached almost the same efficiency for each FHit bin regardless of the zenith angle. The most significant result is the improvement in the $Q$ factor for high zenith angles by a factor of 4 for large showers (high energies). In addition, we present the performance in the reconstruction of showers with cores landing off the array, which are newly available in Pass 5. In comparison to on-array events, the effective area for off-array events is slightly larger and the angular resolution is worse, but the gamma/hadron separation efficiency is better.

We verified the performance of the Pass 5 reconstruction algorithm with the observation of the Crab Nebula with a significance of 258$\sigma$ over 2565 days, using on-array data and 123~$\sigma$ using off-array data. In both cases, we selected events triggering specific PMT ranges to represent small, medium, and large showers. As expected, there are significantly more events at lower energies, but the angular resolution is worse and the significance lower. We fit the Pass 5 log parabola spectral energy distribution for the Crab using our three analytical approaches and found good agreement with other experiments. In addition, we present for the first time an energy dependent study of systematic uncertainties for the FHit analysis. The three analytical approaches show a total systematic uncertainty of 10\% around 1 TeV, and are bounded by $\pm20$\% in the entirety of the energy range. 

Last but not least, using the three analysis approaches, we showed the sensitivity of the improved HAWC reconstruction algorithm for four different declinations: from overhead to 48$^{\circ}$ zenith, where the Galactic Center culminates relative to HAWC. We confirm that the addition of off-array events improves the sensitivity by a factor of 2 at 100 TeV. The sensitivity for GP and NN are similar and exhibit flux thresholds smaller than 10\% of the Crab flux between 400 GeV and 100 TeV at 4$^{\circ}$ and 19$^{\circ}$ declination. However, for Crab-like point sources located up to 30$^{\circ}$ zenith, the sensitivity of HAWC is $\lesssim$~10\% of the Crab Nebula's flux between 2 and 50 TeV with all three analysis approaches.

Thanks to these enhancements, HAWC can now detect sources above the 5$\sigma$ significance threshold previously invisible to us, like the Galactic Center, motivating the preparation of an updated source catalog.

\begin{acknowledgments}

We acknowledge the support from: the US National Science Foundation (NSF); the US Department of Energy Office of High-Energy Physics; the Laboratory Directed Research and Development (LDRD) program of Los Alamos National Laboratory; Consejo Nacional de Ciencia y Tecnolog\'{i}a (CONACyT), M\'{e}xico, grants 271051, 232656, 260378, 179588, 254964, 258865, 243290, 132197, A1-S-46288, A1-S-22784, CF-2023-I-645, c\'{a}tedras 873, 1563, 341, 323, Red HAWC, M\'{e}xico; DGAPA-UNAM grants IG101323, IN111716-3, IN111419, IA102019, IN106521, IN114924, IN110521 , IN102223; VIEP-BUAP; PIFI 2012, 2013, PROFOCIE 2014, 2015; the University of Wisconsin Alumni Research Foundation; the Institute of Geophysics, Planetary Physics, and Signatures at Los Alamos National Laboratory; Polish Science Centre grant, DEC-2017/27/B/ST9/02272; Coordinaci\'{o}n de la Investigaci\'{o}n Cient\'{i}fica de la Universidad Michoacana; Royal Society - Newton Advanced Fellowship 180385; Generalitat Valenciana, grant CIDEGENT/2018/034; The Program Management Unit for Human Resources \& Institutional Development, Research and Innovation, NXPO (grant number B16F630069); Coordinaci\'{o}n General Acad\'{e}mica e Innovaci\'{o}n (CGAI-UdeG), PRODEP-SEP UDG-CA-499; Institute of Cosmic Ray Research (ICRR), University of Tokyo. H.F. acknowledges support by NASA under award number 80GSFC21M0002. We also acknowledge the significant contributions over many years of Stefan Westerhoff, Gaurang Yodh and Arnulfo Zepeda Dom\'inguez, all deceased members of the HAWC collaboration. Thanks to Scott Delay, Luciano D\'{i}az and Eduardo Murrieta for technical support.
\end{acknowledgments}
%%%%%%%%%%%%%%%%%%%%%%%%%

\bibliography{sample631}{}
\bibliographystyle{aasjournal}



\end{document}